\documentclass{aa}
\usepackage{graphicx}
\usepackage{natbib}
\bibpunct{(}{)}{;}{a}{}{,}

\newcommand{\id}{\mbox{$\mathrm{d^{-1}}$}}
\newcommand{\Msun}{\mbox{$M_{\odot}$}}
\newcommand{\kms}{\mbox{$\mathrm{km\,s^{-1}}$}}
\newcommand{\Mwd}{\mbox{$M_\mathrm{wd}$}}
\newcommand{\Msec}{\mbox{$M_\mathrm{sec}$}}
\newcommand{\Teff}{\mbox{$T_\mathrm{eff}$}}
\newcommand{\Porb}{\mbox{$P_{\rm orb}$}}

\newcommand{\linet}[2]{\Ion{#1}{#2}}
\newcommand{\Line}[3]{\Ion{#1}{#2}\,$\lambda$\,#3}
\newcommand{\Lines}[3]{\Ion{#1}{#2}\,$\lambda\lambda$\,#3}
\newcommand{\Ion}[2]{#1{\,\scriptsize #2}}

\newcommand{\Ha}{\mbox{${\mathrm H\alpha}$}}
\newcommand{\Hb}{\mbox{${\mathrm H\beta}$}}
\newcommand{\Hg}{\mbox{${\mathrm H\gamma}$}}
\newcommand{\Hd}{\mbox{${\mathrm H\delta}$}}

\newcommand{\Kwd}{\mbox{$\mathrm{K}_1$}}
\newlength{\width}
\newcommand{\FN}[1]{\settowidth{\width}{$^{(#1)}$}\mbox{$^{(#1)}$}\hspace*{-\width}}
\newcommand{\fn}[1]{\mbox{$^{(#1)}$}}

\begin{document}

\title{HS\,2331+3905: The cataclysmic variable that has it
all\thanks{Based in part on observations made with the NASA/ESA Hubble
Space Telescope, obtained at the Space Telescope Science Institute,
which is operated by the Association of Universities for Research in
Astronomy, Inc., under NASA contract NAS 5-26555, on observations made
at the 1.2m telescope, located at Kryoneri Korinthias, and owned by
the National Observatory of Athens, Greece, and with the Isaac Newton
Telescope and William Herschel Telescope, which are operated on the
island of La Palma by the Isaac Newton Group in the Spanish
Observatorio del Roque de los Muchachos of the Instituto de
Astrofisica de Canarias.}}

\author{
S. Araujo-Betancor\inst{1}\thanks{Visiting Astronomer,
German-Spanish Astronomical Center, Calar Alto, operated by the
Max-Planck-Institut f\"{u}r Astronomie, Heidelberg, jointly with the
Spanish National Commission for Astronomy.} 
   \and
B.T. G\"ansicke\inst{2}$^{\star\star}$ 
   \and 
H.-J. Hagen\inst{3} 
   \and 
T.R. Marsh\inst{2} 
   \and \\
E.T. Harlaftis\inst{4} 
   \and
J. Thorstensen\inst{5} 
   \and 
R.E. Fried\inst{6} 
   \and
P. Schmeer\inst{7}
   \and 
D. Engels\inst{3}
}
 
\offprints{S. Araujo-Betancor, e-mail: araujo@stsci.edu}

\institute{
Space Telescope Science Institute, 3700 San Martin Drive,
Baltimore, MD\,21218, USA
  \and
Department of Physics, University of Warwick, Coventry CV4 7AL, UK
  \and
Hamburger Sternwarte, Universit\"at Hamburg, Gojenbergsweg 112,
21029 Hamburg, Germany
  \and
Institute of Space Applications and Remote Sensing, National
Observatory of Athens, P.O. Box 20048, Athens 11810, Greece
  \and
Department of Physics and Astronomy, Dartmouth College, 6127 Wilder
Laboratory, Hanover, NH 03755-3528, USA
  \and
Braeside Observatory, PO Box 906, Flagstaff AZ 86002, USA
  \and
Bischmisheim, Am Probstbaum 10, 66132 Saarbr\"ucken, Germany
}

\date{Received \underline{\hskip2cm} ; accepted \underline{\hskip2cm} }

\abstract{We report detailed follow-up observations of the cataclysmic
  variable HS\,2331+3905, identified as an emission-line object in the
  Hamburg Quasar Survey. An orbital period of 81.08\,min is
  unambiguously determined from the detection of eclipses in the light
  curves of HS\,2331+3905. A second photometric period is consistently
  detected at $P \simeq 83.38$\,min, $\sim 2.8\,\%$ longer than \Porb,
  which we tentatively relate to the presence of permanent superhumps.
  High time resolution photometry exhibits short-timescale variability
  on time scales of $\simeq5-6$\,min which we interpret as non-radial
  white dwarf pulsations, as well as a coherent signal at 1.12\,min,
  which is likely to be the white dwarf spin period. A large-amplitude
  quasi-sinusoidal radial velocity modulation of the Balmer and Helium
  lines with a period $\sim 3.5$\,h is persistently detected
  throughout three seasons of time-resolved spectroscopy. However,
  this spectroscopic period, which is in no way related to the orbital
  period, is not strictly coherent but drifts in period and/or phase
  on time scales of a few days. Modeling the far-ultraviolet to
  infrared spectral energy distribution of HS\,2331+3905, we determine
  a white dwarf temperature of $\Teff\simeq10\,500$\,K (assuming
  $M_{\rm wd}=0.6$\,\Msun), close to the ZZ\,Ceti instability strip of
  single white dwarfs. The spectral model implies a distance of
  $d=90\pm15$\,pc, and a low value for the distance is supported by
  the large proper motion of the system,
  $\mu=0.14\arcsec\,\mathrm{yr^{-1}}$. The non-detection of molecular
  bands and the low $J$, $H$, and $K$ fluxes of HS\,2331+3905 make
  this object a very likely candidate for a brown-dwarf donor.
   }

\maketitle

\section{Introduction}
In the current standard model of cataclysmic variable (CV) evolution
\citep[see ][]{king88-1} the systems evolve to shorter orbital periods,
reach a minimum period at $\sim 70$\,min and then begin to evolve back
to longer orbital periods. As a CV approaches the minimum period, the
mass of the donor star decreases until it can no longer sustain
hydrogen burning. The low mass secondary star becomes increasingly
degenerate, entering the brown dwarf regime. \citet{kolb93-1}
estimated that 70\% of all CVs should have such ``substellar''
secondaries. Despite these large predicted numbers, only a handful of
CVs are currently believed to harbour a brown dwarf donor, with
WZ\,Sge, EG\,Cnc and EF\,Eri being among the best candidates
\citep{littlefairetal03-1, patterson98-1, beuermannetal00-1}. So far,
for none of them does a compelling proof of the degenerate nature of
the star exist. The discrepancy between the observed and predicted
numbers of brown dwarf CVs could be due to selection effects: because
of the low mass transfer rates predicted for these systems they may be
substantially fainter than most CVs and may have very rare outbursts.
Alternatively, it may also be that the standard theory is wrong , and
that CVs have either not reached the period minimum where their donors
become degenerate \citep{king+schenker02-1}, or they may even die
before reaching the period minimum \citep{patterson98-1}.

In order to be able to  test properly any of these hypotheses we need
to reduce the observational biases that afflict the current CV
population to a minimum. This is the main purpose of an undergoing
large-scale search for CVs using the Hamburg Quasar Survey
\citep[HQS;][]{hagenetal95-1, gaensickeetal00-1, gaensickeetal02-2,
  gaensickeetal04-1, araujo-betancoretal03-2, rodriguez-giletal04-1}. One of
our most recent discoveries, HS\,2331+3905 (HS\,2331 thereafter), was
selected as a CV candidate on the basis of its spectral
characteristics in the HQS. Specifically, its HQS spectrum displays the strong Balmer
emission lines that suggest ongoing accretion in
CVs. 

In this paper we present the results of follow-up spectroscopy and
photometry of HS\,2331 obtained over three years, which confirm the CV
nature of this object and unveil a very unusual picture.  In
Sec.\,\ref{s-obs} we describe the photometric and spectroscopic
(ground and space) observations and data reduction and calibration.
The analysis of the photometry and ground-based spectroscopy is
presented in Sec.\,\ref{s-ana_photo} and \ref{s-ana_spec}. In
Sec.\,\ref{s-ana_spec} we also present a model for the spectral energy
distribution of HS\,2331, from which we derive a temperature for the
white dwarf and distance to the system.  Finally, in
Sec.\,\ref{s-discussion}, we use our findings in trying to construct a
preliminary picture of the nature of HS\,2331.

\section{Observations and Data Reduction}
\label{s-obs} 
\begin{table*}[t]
\caption[]{\label{t-obslog}Log of Observations.}
\setlength{\tabcolsep}{1.2ex}
\begin{minipage}[t]{8.8cm}

\begin{flushleft}
\begin{tabular}[t]{lcccc}
\hline\noalign{\smallskip}
\multicolumn{5}{c}{Photometry}\\
Date & UT Time &  Filter & Exp.\,[s] & \#\,Frames \\
\noalign{\smallskip}\hline
\noalign{\smallskip}
\multicolumn{5}{l}{\textbf{40\,cm Braeside Observatory}}\\
2000 Sep 25 & 04:35 - 12:29 &  $R$     & 50 & 510 \\
2000 Sep 28 & 05:26 - 10:54 &  $R$     & 85 & 211 \\
2000 Oct 03 & 02:21 - 12:01 &  $R$     & 95 & 338 \\
2001 Nov 21 & 04:13 - 08:44 &  C     & 85 & 152 \\
2001 Nov 22 & 02:43 - 08:45 &  C     & 55 & 302 \\
2003 Oct 12 & 02:19 - 11:44 &  C     & 55 & 552 \\
2003 Oct 13 & 02:36 - 11:30 &  C     & 55 & 542 \\
2003 Oct 15 & 03:56 - 11:23 &  C     & 55 & 446 \\
\noalign{\smallskip}
\multicolumn{5}{l}{\textbf{1.2\,m Kryoneri Observatory}}\\
2002 Oct 15 & 18:41 - 22:39 &  C     & 30  & 340 \\
2002 Oct 16 & 17:29 - 22:13 &  C     & 25  & 429 \\
2003 Aug 14 & 22:56 - 02:52 &  C     & 10  & 1029 \\
2003 Aug 15 & 23:14 - 02:57 &  C     & 10  & 1012 \\
2003 Aug 16 & 23:33 - 03:06 &  C     & 10  & 980 \\
2003 Aug 17 & 23:32 - 03:04 &  C     & 10  & 948 \\
2003 Aug 18 & 22:23 - 03:01 &  C     & 10  & 1249 \\
2003 Aug 19 & 22:09 - 03:04 &  C     & 10  & 1311 \\
2003 Aug 20 & 21:56 - 03:07 &  C     & 10  & 1432 \\
\noalign{\medskip}
\multicolumn{5}{l}{\textbf{4\,m WHT/ULTRACAM}}\\
2003 Nov 10 & 19:23 - 22:00 & $u'$    & 1   & 10013 \\
2003 Nov 10 & 19:23 - 22:00 & $g'$    & 1   & 10013 \\
2003 Nov 10 & 19:23 - 22:00 & $r'$    & 1   & 10013 \\
\noalign{\smallskip}\hline
\end{tabular}
\end{flushleft}
\end{minipage}
\hfill
\begin{minipage}[t]{8.8cm}
\begin{flushleft}
\begin{tabular}[t]{lcccc}
\hline\noalign{\smallskip}
\multicolumn{5}{c}{Spectroscopy}\\
Date & UT Time &  Grating & Exp.\,[s] & \#\,Spectra \\
\noalign{\smallskip}\hline
\noalign{\smallskip}
\multicolumn{5}{l}{\textbf{2.2\,m Calar Alto/CAFOS}}\\
2000 Sep 20    &  00:50         & B-200    & 600 & 1 \\
2000 Sep 20    &  01:05         & R-200    & 600 & 1 \\
2000 Sep 23    &  21:20 - 02:30 & B-100    & 600 & 24\\
2000 Sep 24    &  00:22 - 03:37 & B-100    & 600 & 16\\
2000 Sep 24    &  03:47 - 04:27 & R-100    & 600 & 3 \\
2003 Sep 11    &  22:16 - 04:20 & G-100    & 600 & 28\\
2003 Sep 15    &  02:28 - 04:35 & G-100    & 600 & 12\\
2003 Sep 16    &  23:48 - 04:04 & G-100    & 600 & 23\\
\noalign{\smallskip}
\multicolumn{5}{l}{\textbf{2.5\,m Isaac Newton Telescope/IDS}}\\
2002 Aug 26    &  02:37 - 03:60 & R632V    & 600 & 9\\
2002 Aug 28    &  00:33 - 01:56 & R632V    & 600 & 9\\
2002 Sep 01    &  04:16 - 04:47 & R632V    & 600 & 4\\
2002 Sep 02    &  03:09 - 03:41 & R632V    & 600 & 4\\
\noalign{\smallskip}
\multicolumn{5}{l}{\textbf{Hubble Space Telescope/STIS}}\\
2002 Oct 24    &  11:34         & G140L    & 700 & 1 \\
\noalign{\smallskip}
\multicolumn{5}{l}{\textbf{2.4\,m MDM Hiltner}}\\
2003 Oct 12    &  07:28 - 08:15  & 600 & 420 & 7  \\
2003 Oct 13    &  07:38 - 08:17  & 600 & 420 & 6  \\
2003 Oct 14    &  08:03 - 08:59  & 600 & 420 & 8  \\
2003 Oct 15    &  02:51 - 10:48 & 600 & 420 & 21 \\
2003 Oct 16    &  03:45 - 04:40  & 600 & 420 & 8  \\
\noalign{\smallskip}\hline
\end{tabular}
\end{flushleft}
\end{minipage}
\end{table*}

\begin{figure}
\centerline{\includegraphics[width=7cm]{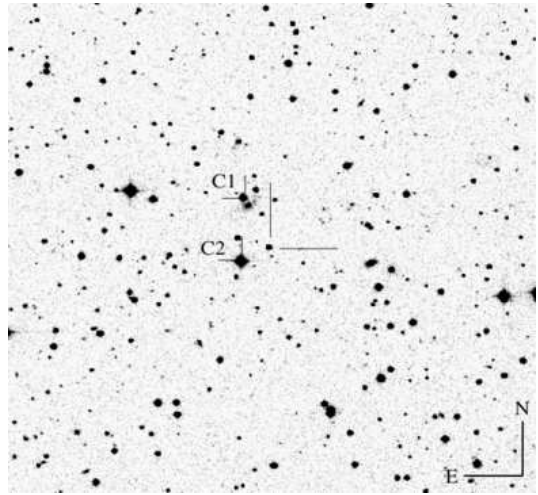}}
\caption[]{\label{f-fc} $10\arcmin\times10\arcmin$ Finding chart of
HS\,2331, obtained from the Digitized Sky Survey~2. The coordinates of
the CV are $\alpha(2000)=23^h34^m01.6^s$,
$\delta(2000)=+39^{\circ}21\arcmin41.4\arcsec$. Differential
magnitudes of HS\,2331 were obtained relative to the comparison stars
C1 (USNO--A2.0~1275--18486676) and C2 (GSC\,0323100595).}
\end{figure}

\subsection{\label{s-obs_photo} Photometry}

\begin{figure*}
\includegraphics[width=8.8cm]{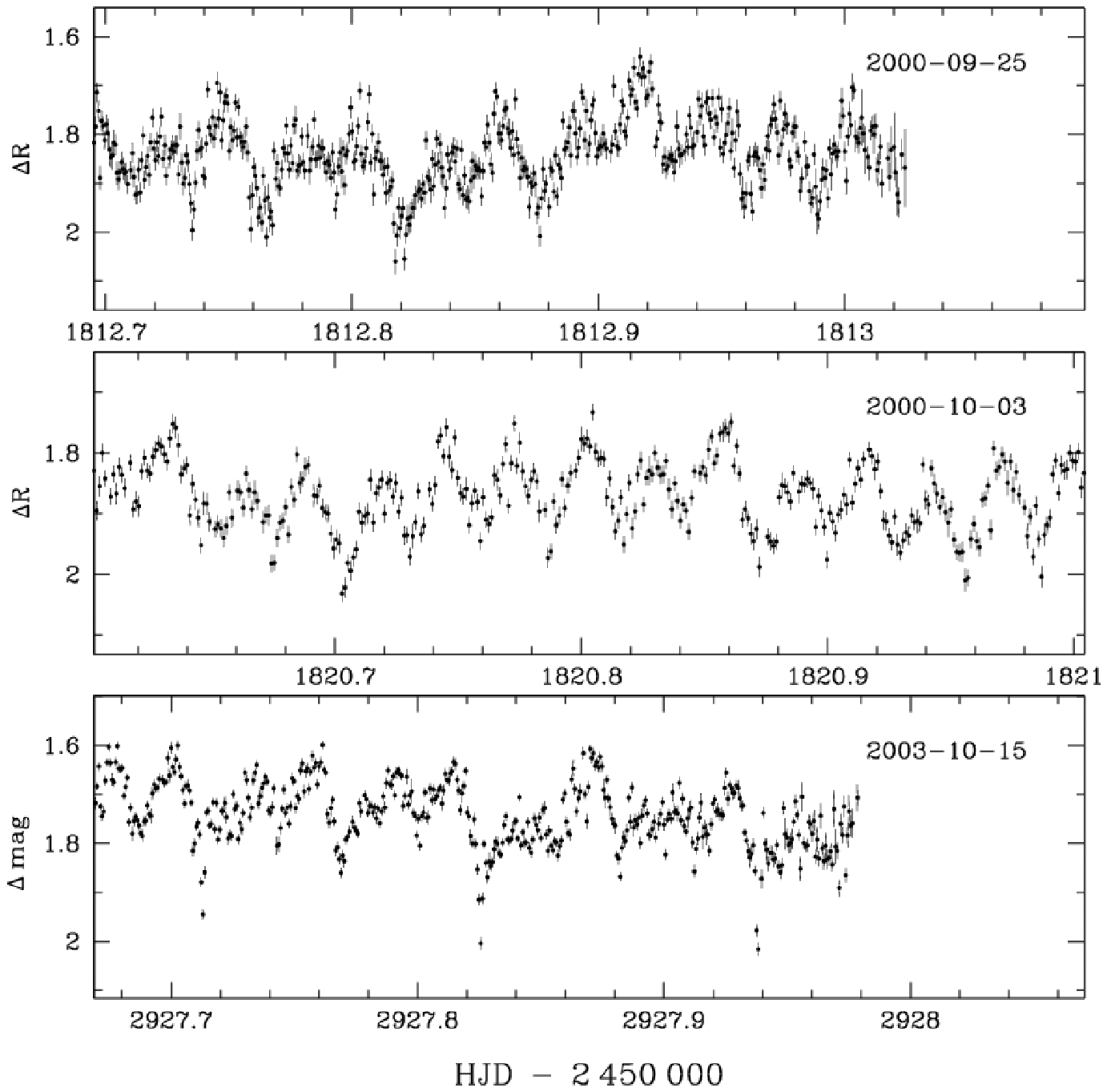}
\includegraphics[width=8.8cm]{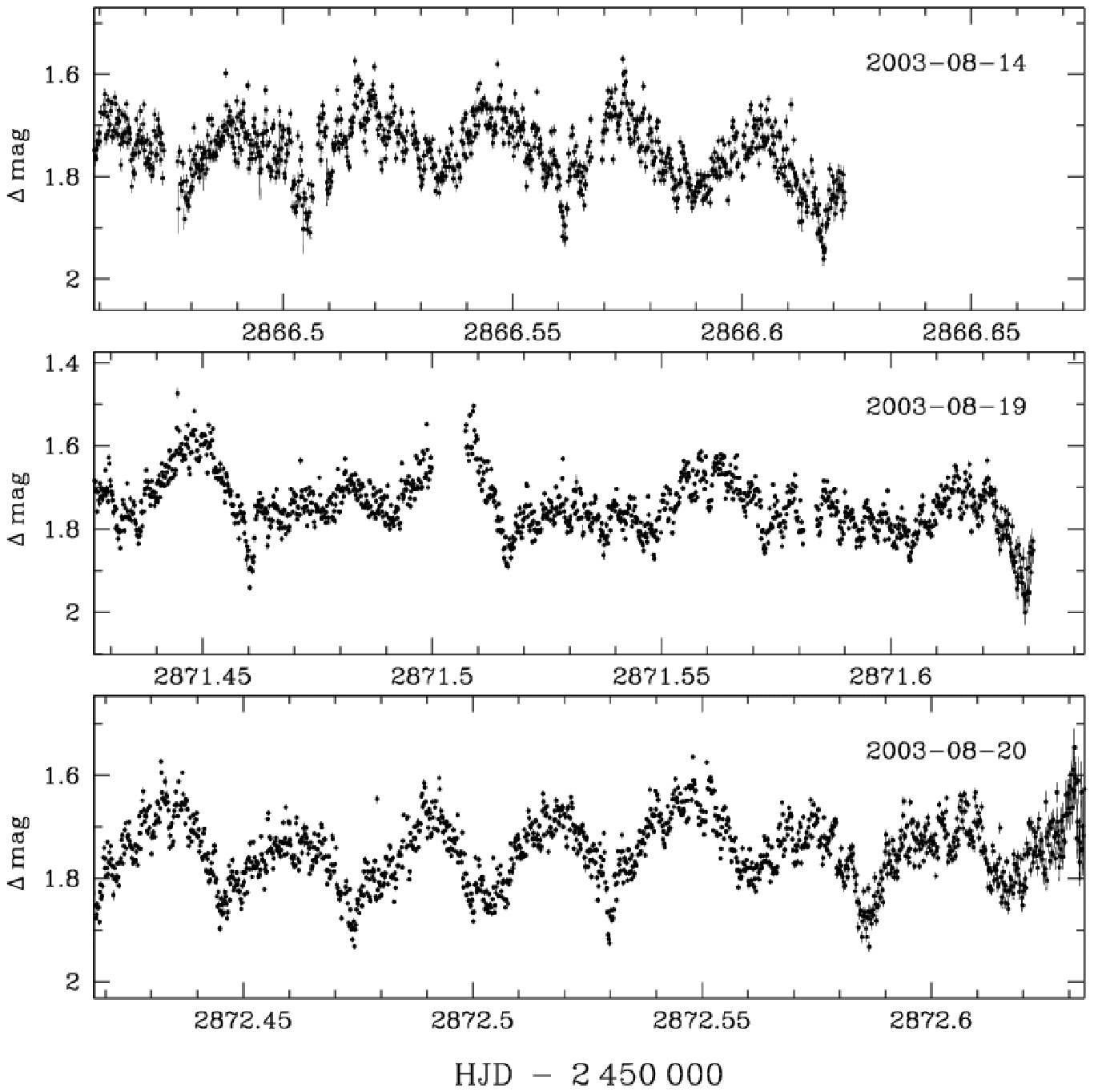}

\caption[]{\label{f-lc_braeside_kryoneri} Sample light curves of HS\,2331
based on  differential CCD photometry obtained at the 40\,cm Braeside
Observatory (left) and at the 1.2\,m Kryoneri
Observatory (right).}

\includegraphics[angle=0,width=18cm]{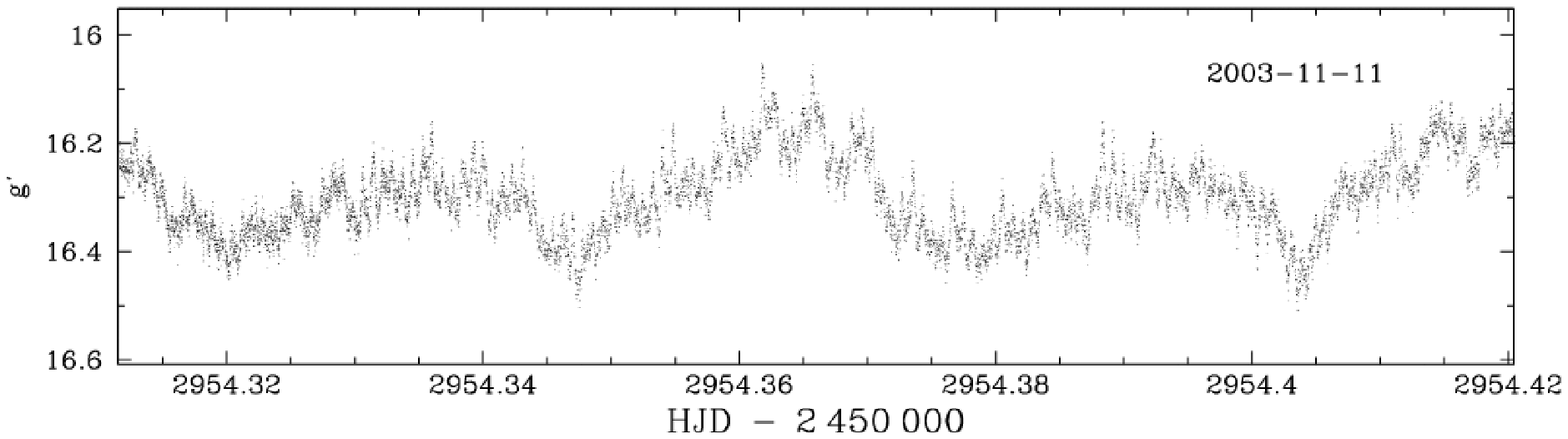}
\caption[]{\label{f-lc_ultracam} Light curve of HS\,2331 obtained at a
time resolution of 1\,s with the $g'$-filter using ULTRACAM on the
WHT.}

\end{figure*}

\subsubsection{Braeside Observatory}
Differential $R$-band CCD photometry of HS\,2331 was obtained at the
Braeside Observatory immediately following the identification of its
CV nature in September-October 2000 (Sect.\,\ref{s-calaralto}), using
a 0.4\,m reflector equipped with a SITe\,512 CCD camera
(Table\,\ref{t-obslog}). HS\,2331 was again observed from Braeside in
November 2001 and October 2003, on both occasions in white light. The
data were bias-subtracted, dark current-subtracted and flat-fielded in
a standard fashion, and instrumental aperture magnitudes of HS\,2331
were derived relative to USNO--A2.0~1275--18486676 ($R=14.6$, labelled
`C1' in Fig.\,\ref{f-fc}). The mean magnitude of HS\,2331 during the
2000 observations was $R\simeq16.5$.

The Braeside CCD light curves of HS\,2331 display periodic variability
with an amplitude of $\sim\pm0.1$\,mag
(Fig.\,\ref{f-lc_braeside_kryoneri}), which has consistently been
detected in all three years covered by our observations.  The
morphology of the photometric modulation is best described by a
double-humped pattern with a period of $\sim80$\,min. The relative
strength of the two humps varies substantially between the individual
nights. The 2003 data reveal narrow absorption dips centred on some of
the observed minima between the humps.

\subsubsection{Kryoneri Observatory}
Filterless CCD photometry of HS\,2331 was obtained during two nights
in October 2002 and during 7 nights in August 2003 at the 1.2\,m
Kryoneri telescope using a SI-502 $516\times516$ CCD camera
(Table\,\ref{t-obslog}). The Kryoneri data were reduced following the
procedure described in \citet{gaensickeetal04-1}, and differential
magnitudes were derived relative to C1 (Fig.\,\ref{f-fc}). 

The October 2002 light curves, obtained with a time resolution of
25\,s -- 30\,s, were essentially similar to the Braeside observations
(Fig.\,\ref{f-lc_braeside_kryoneri}), displaying a double-humped structure with
a period of $\sim80$\,min. The higher time resolution data obtained in
August 2003 provide stronger evidence for the absorption dips detected
in the Braeside observations, and clearly reveal additional short-term
variability on a time scale of $\sim5$\,min (e.g. August 14;
Fig.\,\ref{f-lc_braeside_kryoneri}). In addition the changes of the shape of
the double-humped structure are more pronounced than in the Braeside
observations. On August 19, one of the humps is extremely weak, on
August 20 both humps have nearly equal amplitude. 
 
\subsubsection{ULTRACAM on the William Herschel Telescope}
The highest time resolution and signal-to-noise ratio CCD photometry
was obtained on November 10, 2003 using the 3-beam ULTRACAM CCD camera
\citep[which takes CCD data in three colour channels simultaneously,
][]{dhillonetal02-1} on the William Herschel Telescope. HS\,2331 was
observed with a 1\,s time resolution, using Sloan $u'$, $g'$, and $r'$
filters. The data were reduced with the ULTRACAM
pipeline. Differential magnitudes of HS2331 were obtained relative to
GSC\,0323100595 (labelled `C2' in Fig.\,\ref{f-fc}), and converted to
apparent magnitudes using observations of the Sloan standard
SA100-280, which was observed at the end of the night. The mean
magnitudes of HS2331 were $u'=16.1$, $g'=16.3$ and $r'=16.0$. These
values are subject to some small systematic uncertainty related to the
yet uncalibrated colour terms accounting for the differences between
the ULTRACAM and Sloan setups.

The ULTRACAM light curves impressively confirm the presence of
variability on a time scale of $\sim5$\,min seen in the Kryoneri data,
as well as additional variability on even shorter time scales
(Fig.\,\ref{f-lc_ultracam}). 

\subsection{Optical Spectroscopy}
\subsubsection{Calar Alto\label{s-calaralto}}

HS\,2331 was identified as a CV from low-resolution spectroscopy
obtained at the Calar Alto 2.2\,m telescope on September 20 2000. The
CAFOS focal reductor spectrograph was used in conjunction with the
standard SITe CCD. A single pair of blue/red identification spectra
was obtained with the B-200 and R-200 grating through a $2\arcsec$
slit  (see Table\,\ref{t-obslog}). The wavelength range of the
combined spectra extends from 3500 to 10000\,\AA\ at a spectral
resolution of approximately 10\,\AA. The online reduction of the
identification spectra using the CAFOS quick look context within MIDAS
immediately revealed the CV nature of HS\,2331 by the presence of broad
double peaked Balmer and He\,I emission lines in the spectra.

Time-resolved follow-up spectroscopy of HS\,2331 was carried out
during the same observing run, on September 23 and 24 2000. A total of
43 spectra of 600\,s each were obtained during the two nights, which
were mainly photometric except for very thin cirrus during the last
$\sim1.5$\,hours of the second night (Table\,\ref{t-obslog}). The
B-100 grism was used along with a $1.5\,\arcsec$ slit to obtain 40
spectra with a useful wavelength range of 3500--6300\,\AA\ and a
spectral resolution of $\sim4$\,\AA. In addition, 3 red spectra were
acquired during the second night, using the R-100 grism with a
$1.5\,\arcsec$ slit width. The red spectra cover a wavelength range of
6000 to 9200 \AA\ at a spectral resolution of $\sim 4$\,\AA.
Mercury-cadmium, helium-argon and rubidium arcs were taken regularly
throughout both nights to provide for the wavelength calibration.
Data from the flux standards G191B2B and $\zeta$\,Cas were also
obtained to correct for the instrument response of the ``blue" and
``red" spectra respectively. The CAFOS acquisition image were taken
with a Johnson $V$ filter in order to correct the spectra for slit
losses.

All spectra were reduced in a standard manner using the
\texttt{Figaro} package within the Starlink software collection. The
bias level was subtracted from the calibration and object images using
the median from a set of unexposed frames. Dome flat field images,
taken with the same set up as the object frames, were used to remove
pixel to pixel variations. The spectra were then optimally extracted
\citep{horne86-1} and sky line subtracted. A fourth-order polynomial
was fitted to obtain the dispersion relation for each of the arc
spectra, giving an RMS smaller than 0.1 \AA\ in all cases. Each target
spectrum was then wavelength-calibrated by interpolating two
neighboring arc spectra to account for the possible drift of the
wavelength-to-pixel scale. The instrumental response function was
computed as the ratio of a spline fit to the line-free continuum of
the B-100/R-100 spectra of the flux standards to the flux-calibrated
data from \citet{filippenkoetal84-1} and \citet{oke90-1}. All target
spectra were flux-calibrated and corrected for atmospheric extinction
using this response function.

\begin{figure*}
\includegraphics[angle=270, width=8.8cm]{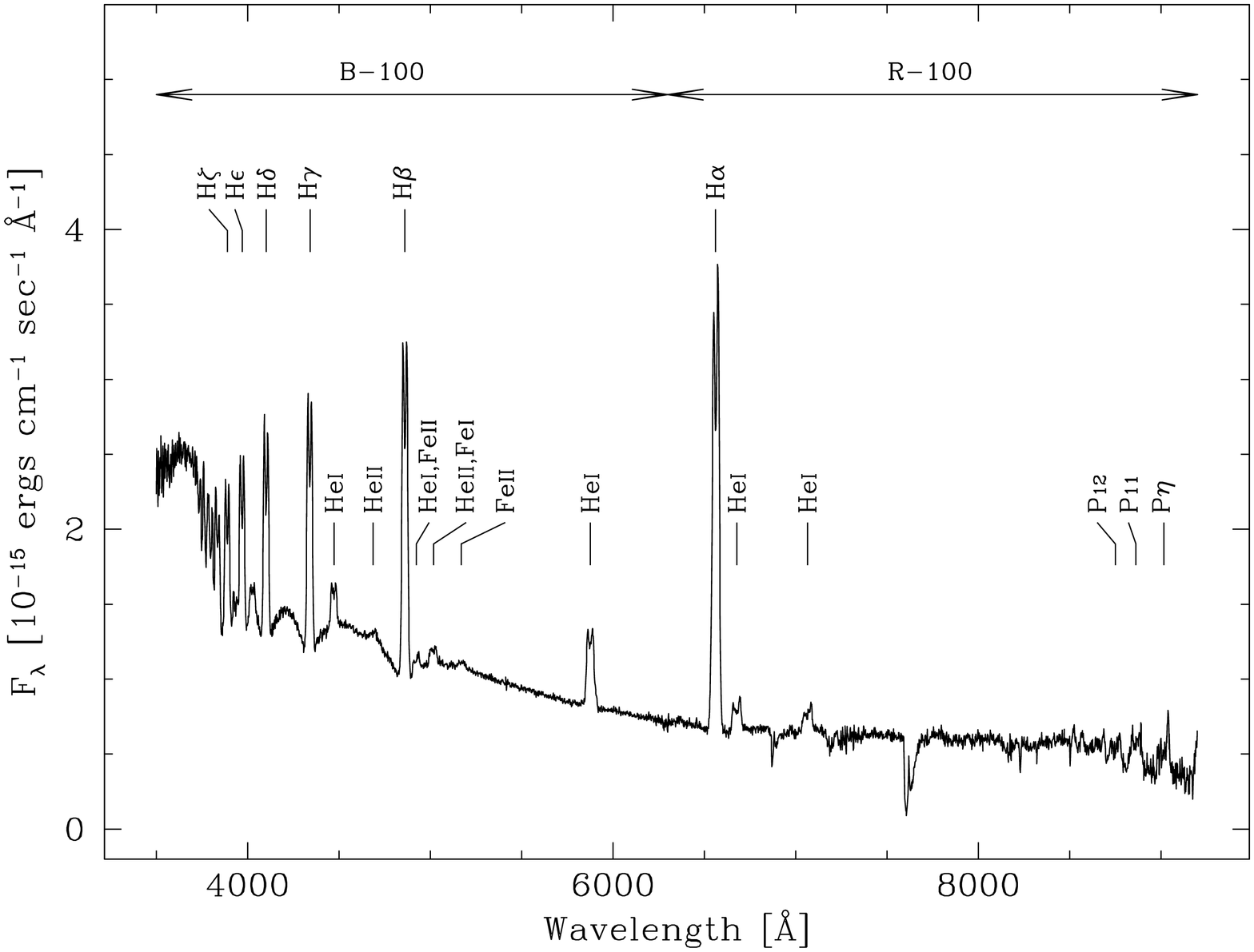}
\includegraphics[angle=270, width=8.8cm]{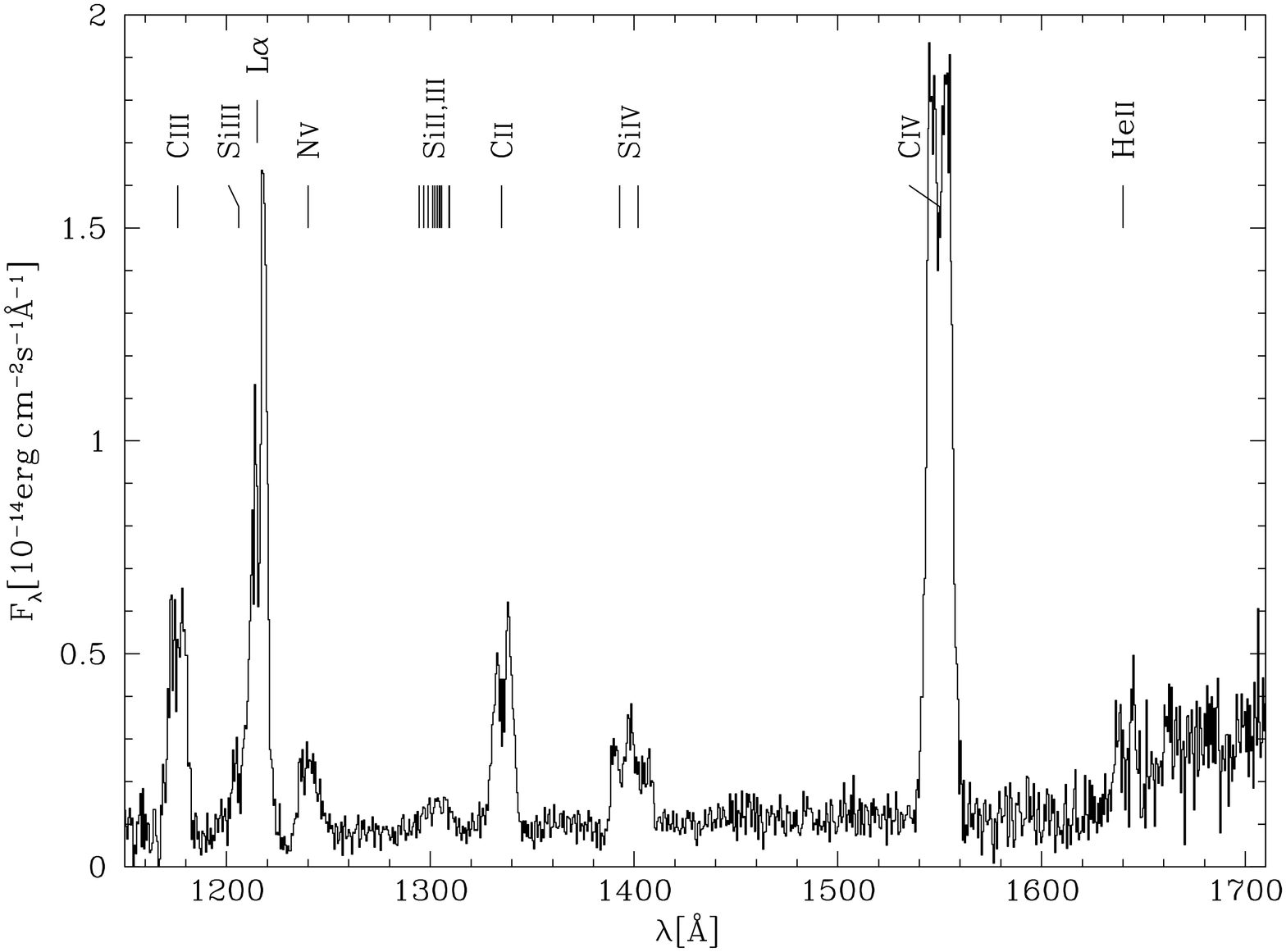}

\begin{minipage}[t]{8.8cm}
\caption[]{\label{f-average} Mean, flux-calibrated ``blue" and ``red"
spectra (Calar Alto 2000, B-100/R-100 data set) of HS\,2331 combined
together.  The arrows indicate the wavelength range covered by the
B-100 and R-100 gratings.}
\end{minipage}
\hfill
\begin{minipage}[t]{8.8cm}
\caption[]{\label{f-hst}  HST/STIS spectrum of HS\,2331.}
\end{minipage}
\end{figure*}

We used the acquisition images obtained in conjunction with the CAFOS
spectroscopy to estimate the brightness of HS\,2331 during our Calar
Alto observations. Aperture photometry relative to GSC\,0323100595
($V=12.75$) located $\sim35\,\arcsec$ south-east of HS\,2331 (`C2' in
Fig.\,\ref{f-fc}) resulted in $V\simeq16.4$, which is consistent with
the magnitude obtained from the HQS direct plate, and with the
brightness level of all our photometry (Sect.\,\ref{s-obs_photo}). A
grand average spectrum of HS\,2331 was created by combining the mean
of all ``blue'' and ``red'' spectra from the flux-calibrated (and
corrected for atmospheric extinction) B-100/R-100 data set. This
average spectrum was then scaled to $V=16.4$, as determined from the
 CAFOS acquisition images to account for possible slit losses
(Fig.\,\ref{f-average}). The applied correction of $\simeq0.2$\,mag is
comparable to the intrinsic variability of HS\,2331, and indicates
that the absolute error in our flux calibration is $\le20$\,\%.

A further set of observations were obtained in September 2003 at the
2.2\,m Calar Alto telescope.  This time the G-100 grism was used which
together with a slit width of $1.2\arcsec$ provided a wavelength range
of 4240--8300\,\AA\ and a spectral resolution of $\sim 4.1$\,\AA\
(Table\,\ref{t-obslog}). These spectra were reduced in an analogous
fashion as described above.

The spectrum of HS\,2331 (Fig.\,\ref{f-average}) displays broad
double-peaked Balmer and \Ion{He}{I} emission lines, suggesting an
origin in the accretion disc. The higher series of the Balmer lines
are flanked by extremely broad absorption troughs, which are
reminiscent of the Stark-broadened absorption lines observed in the
high-gravity atmospheres of cool white dwarfs (the width of the
absorption lines are too wide to be of accretion disc origin). The red
part of the spectrum does not contain any spectral features that could
be ascribed to the emission of the secondary star.

Table\,\ref{t-ew} lists the equivalent widths (EW) and full width at
half maximum (FWHM, corrected for the instrumental response), of
\Ha--\Hd. The EW and the FWHM
measurements were obtained by fitting a single Gaussian in all
cases. Average spectra were used for the Balmer lines: Calar Alto
2000 for \Hd\ and Calar 2003 for \Ha--\Hd.

\begin{table}[t]
\caption[]{\label{t-ew} EW and FWHM (corrected for the instrumental
resolution) of the strongest Balmer emission lines and of the
strongest FUV lines.} 
\begin{center}
\begin{tabular}{lrcccc}
\hline\noalign{\smallskip}
Line & EW [\AA]  & FWHM [km/s] & FWHM[\AA] \\  
\hline\noalign{\smallskip}
\Ha     &               $185\pm14$  &  $1312\pm5$ & $28.7\pm0.1$\\ 
\Hb     &               $75\pm9$    &  $1475\pm6$ &  $23.9\pm0.1$ \\
\Hg     &               $40\pm4$    &  $1583\pm7$ & $22.9\pm0.1$ \\
\Hd     &                $30\pm13$  &  $1403\pm7$ & $19.2\pm1.1$ \\
\linet{C}{IV}  & $234\pm14$  &  $2478\pm19$ & $12.8\pm0.1$ \\
\linet{Si}{IV} & $52\pm13$ & $3749\pm236$ & $17.5\pm1.1$ \\
\linet{C}{II} & $67\pm12$ & $2476\pm68$ & $11.0\pm0.3$\\
\linet{C}{III} & $111\pm19$ & $2500\pm51$ & $9.8\pm0.2$\\
\linet{N}{V} &   $42\pm15$  &  $2417\pm48$  & $10.0\pm0.2$\\
\noalign{\smallskip}\hline
\end{tabular}
\end{center}
\end{table}

\subsubsection{2.5\,m Isaac Newton Telescope}

A total of 26 spectra of 600\,s each, spread out over a week, were
obtained in August-September 2002 (Table\,\ref{t-obslog}).  We used
the Intermediate Dispersion Spectrograph (IDS) on the 2.5\,m Isaac
Newton Telescope (INT), at the Roche de los Muchachos observatory in
La Palma.  The IDS was equipped with the R632V grating and the $2048
\times 4100$ pixel EEV10 detector. Using a slit width of $1.5\arcsec$
the setup provided an unvignetted wavelength range of
$\sim4400\,\mbox{\AA}-6800\,\mbox{\AA}$ and a spectral resolution of
$\sim2.3$\,\AA. Copper-Argon wavelength calibrations (arcs) were
obtained at the beginning and end of each observation block of
HS\,2331. The reduction procedure applied to this data set is
identical to the one applied to the Calar Alto/CAFOS data above.

\subsubsection{2.4\,m Hiltner telescope}
Most recently, we observed HS\,2331 from the 2.4\,m Hiltner telescope
at MDM observatory (Table\,\ref{t-obslog}). We obtained a total of 50
spectra of 420\,s each in October 2003. The modular spectrograph
combined with the 600\,line\,$\rm mm^{-1}$ grating and a SITe\,$2048
\times 2048$ CCD detector, yielded a spectral dispersion of
$\sim$2.0\,\AA\,$\rm pixel^{-1}$ from 4000 to 7500\,\AA. The reduction
technique applied to this set of data is as explained in
\citet{thorstensenetal98-1}. 

\subsubsection{Hubble Space Telescope}
Far-ultraviolet ($FUV$) spectroscopy of HS\,2331 was obtained with the
Hubble Space Telescope/Space Telescope Imaging Spectrograph (HST/STIS)
on October 24, 2002 as part of an ongoing snapshot survey of CVs. The
observations were obtained using the G140L grating and the
$52\arcsec\times0.2\arcsec$ aperture, providing a spectral resolution
of $R\approx1000$ over the wavelength range $1150-1710$\,\AA. The data
were pipeline-processed within IRAF using CALSTIS V2.13b. The F28x50LP
magnitude of HS2331 measured from the STIS CCD acquistion image
confirms that the system was at its usual brightness level during the
FUV observations. The STIS spectrum (Fig.\,\ref{f-hst}) contains the
mixture of low and high-ionisation emission lines often observed in
CVs: \Line{C}{III}{1176}, \Line{Si}{III}{1206}, \Lines{N}{V}{1239,43},
\Line{C}{II}{1335}, a broad complex of \Ion{Si}{III} near 1300\,\AA,
\Lines{Si}{IV}{1394,1403}, \Lines{C}{IV}{1548,51} and probably
\Line{He}{II}{1640}. The emission line flux ratios are within the
normal range observed in CVs \citep{maucheetal97-1,gaensickeetal03-1}.
Several of the lines contained in the STIS are double-peaked, again
suggesting an origin in an accretion disc. The EW and FWHM of the
strongest lines in the STIS spectrum were estimated by fitting a
Gaussian to the observed line profiles, and the results are reported
in Table\,\ref{t-ew}. Correcting the FWHM for the separation of the
\Ion{C}{II}, \Ion{C}{IV} and \Ion{Si}{IV} doublets results in line
widths which are comparable to those of the Balmer lines, suggesting a
common origin of the FUV and optical emission lines.

The continuum flux underlying the emission lines is slightly
increasing towards the red, with a sudden upturn in flux at
wavelengths $\lambda\ga1650$\,\AA.

\begin{figure*}
\includegraphics[angle=0,width=18cm]{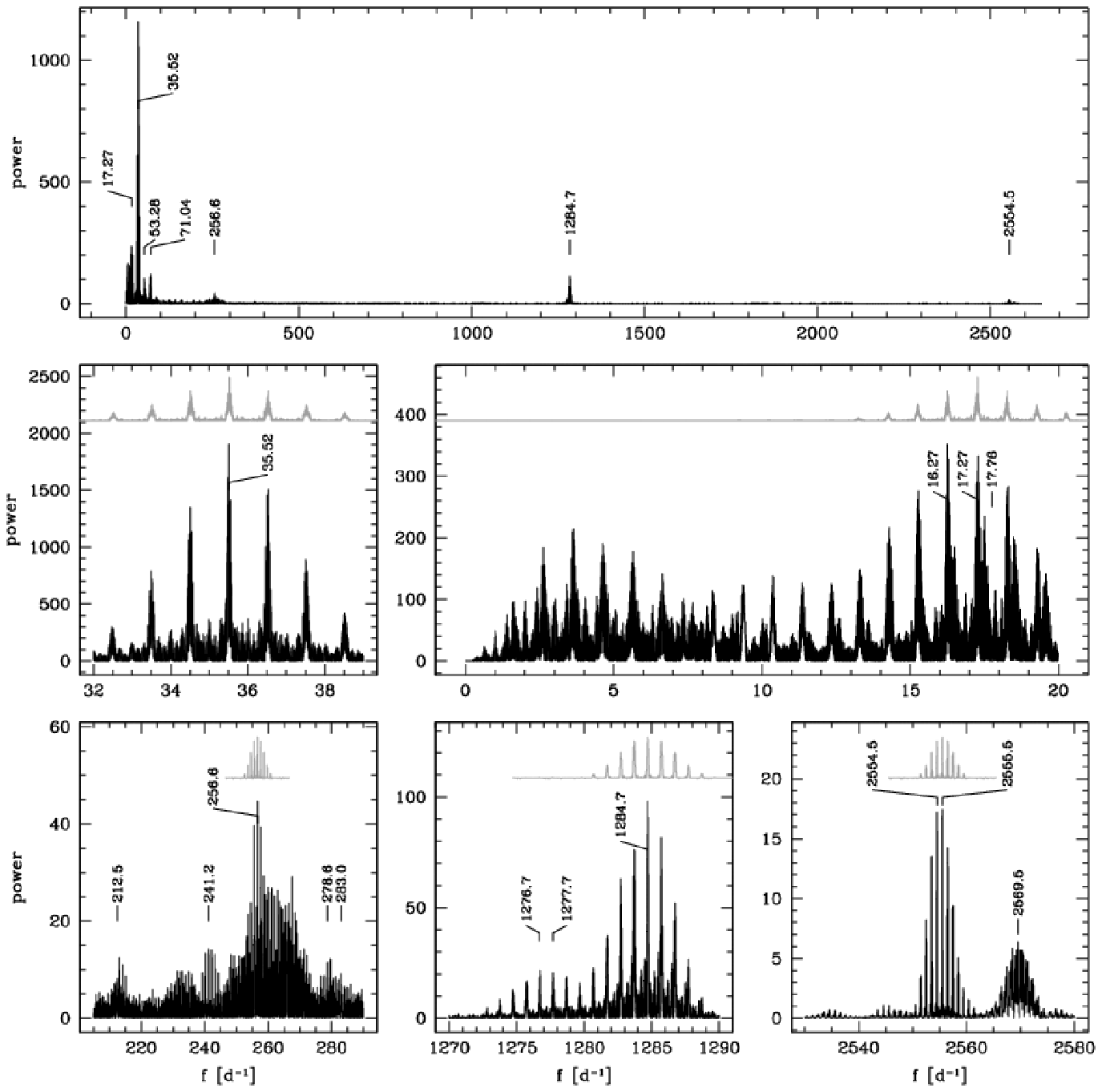}
\caption[]{\label{f-scarglephot} Scargle periodograms computed from
the CCD photometry of HS\,2331. \textbf{Top panel:} The entire
frequency range sampled by the Kryoneri 2003 data, which represents
the largest and best sampled data set to investigate periodicities on
time scales of $\sim2$\,d to $\sim30$\,s. \textbf{Middle panels:}
Close-ups of the power spectra calculated from all available
photometric data (Table\,\ref{t-obslog}). The left panel shows a
coherent signal at 35.52\,\id. Shown in gray is the window function of
the entire data set, shifted to the frequency of the strongest
peak. We interpret this signal as twice the orbital frequency, as the
orbital light curve is dominated by a double-hump structure
(Fig.\,\ref{f-fold81}). The right panel shows several signals in the
low-frequency range. Practically no power is detected at the orbital
frequency (i.e. 17.76\,$\rm d^{-1}$ or 81.08\,min), but a strong
signal is found at 17.27\,\id\ (83.38\,min)  and the one-day alias at
16.27\,\id\ (88.51\,min). \textbf{Bottom panels}: Close-up views of
the high-frequency signals in the Scargle periodogram computed from
the Kryoneri 2003 data only. As in the middle panels, the gray line
shows the alias structure resulting from the sampling of our
observations. Clear signals are detected at 5.61\,min (256.6\,\id),
1.12\,min (1284.7\,\id), and 0.56\,min (2555.5\,\id).}
\end{figure*}

\section{\label{s-ana_photo} Analysis: Photometry}
We have analysed the complex variability detected in the light curves
of HS2331 (for a small sample see Fig.\,\ref{f-lc_braeside_kryoneri}
and \ref{f-lc_ultracam}) by computing Scargle periodograms
\citep{scargle82-1} for the entire CCD photometry of HS\,2331, as well
as for individual sub-sets of the data, using the \texttt{MIDAS/TSA}
context. Prior to this period analysis, we have subtracted the nightly
mean of each light curve to account for night-to-night variations in
the overall brightness of the system and more importantly of the different
detectors/filters used during the observations.  The periodogram of
HS\,2331 calculated from the Kryonery 2003 data, the largest and best
sampled photometric data set, (Fig.\,\ref{f-scarglephot}; top panel)
contains a multitude of signals, as anticipated by our visual
inspection of the light curves (Sect.\,\ref{s-obs_photo}).

The strongest peak is located at 35.52\,$\rm d^{-1}$ (40.54\,min), a
weaker signal is centred on $\sim$17.27\,$\rm d^{-1}$ (83.38\,min),
and some power is also found at 53.28\,$\rm d^{-1}$ (27.03\,min) and
71.04\,$\rm d^{-1}$ (20.27\,min). In the high-frequency range there is
a strong peak at 1284.7\,$\rm d^{-1}$ (1.12\,min) and a weaker one at
256.6\,$\rm d^{-1}$ (5.61 \,min).

\subsection{The orbital period}

Fig.\,\ref{f-scarglephot} shows a close-up view of the Scargle
periodogram of the $\sim$40\,min signal (middle left panel). The
strongest peak is located at $40.542632\pm0.000008$\,min
($35.518168\pm0.000007\,\rm d^{-1}$) with the one $\sigma$ error
obtained from a sine fit to the entire photometric data. The other
peaks seen in the periodogram are aliases separated by multiples of
1\,\id. The sharpness and power of the 40.54\,min signal indicates a
very high degree of coherence, suggesting that it is related to a
stable ``clock'' within the binary system, such as the binary orbital
period or the white dwarf spin period.

\begin{figure*}
\includegraphics[angle=0,width=\columnwidth]{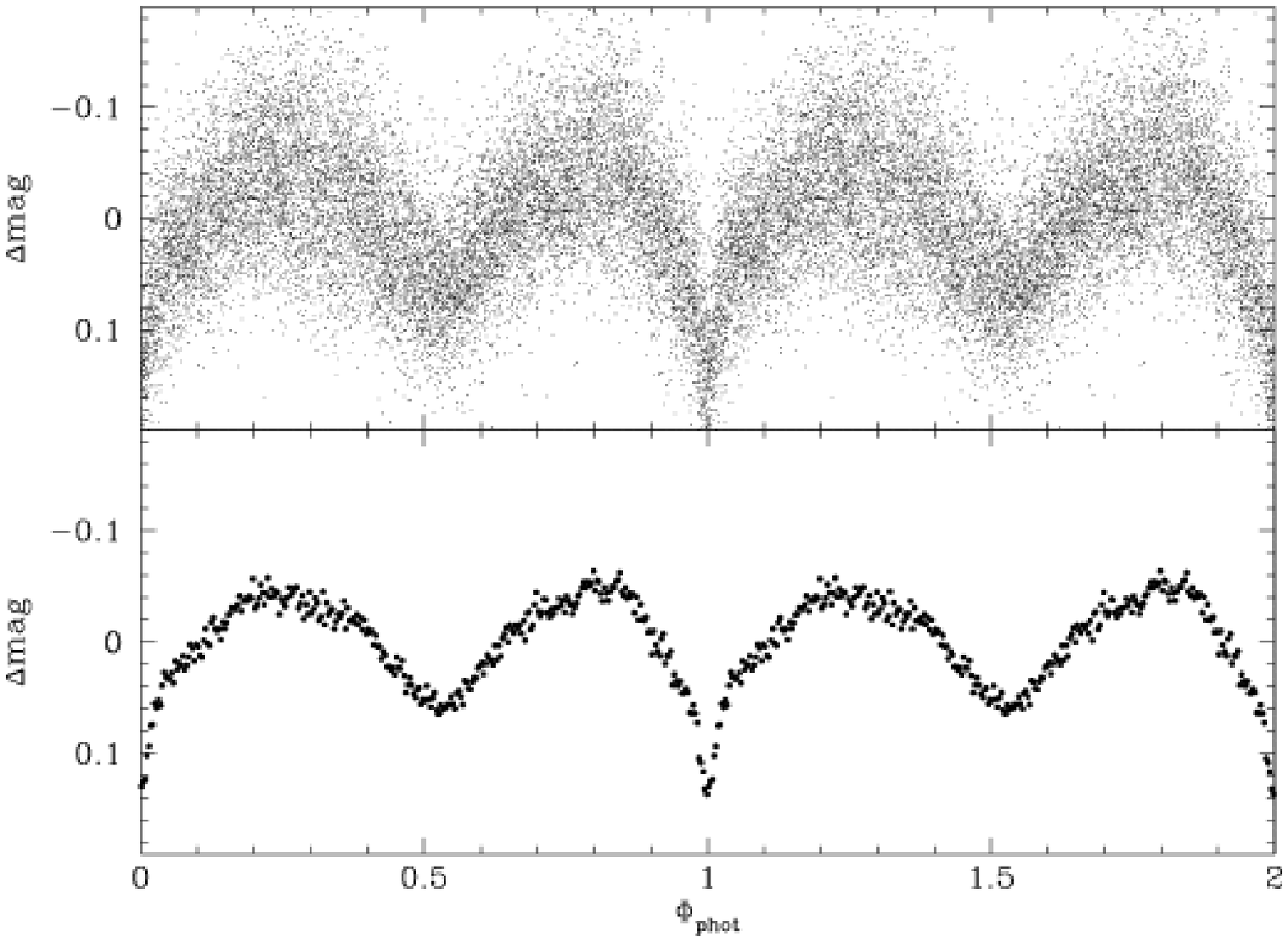}
\hfill
\includegraphics[angle=0,width=\columnwidth]{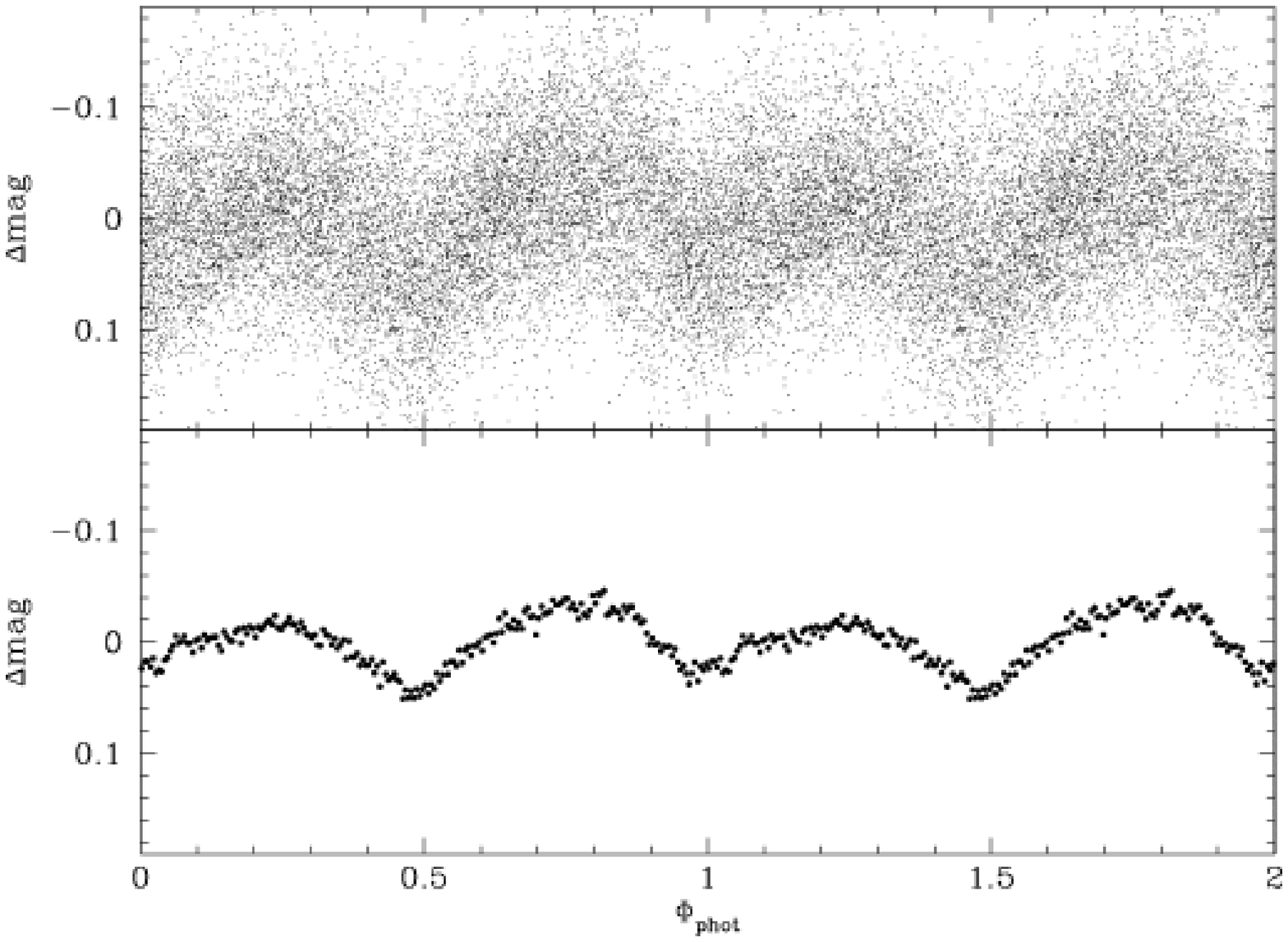}

\parbox{\columnwidth}{\caption{\label{f-fold81} HS\,2331 entire
  photometric data set folded over 81.08\,min (upper panel) and
  re-sampled into bins of 100 points (bottom panel). The light curves
  are repeated over two cycles for clarity.}}
\hfill
\parbox{\columnwidth}{\caption{\label{f-fold83} HS\,2331 entire
    photometric data set folded over 83.38\,min (upper panel) and
    re-sampled into bins of 100 points (bottom panel). The light
    curves are repeated over two cycles for clarity.}}
\end{figure*}

Folding the entire 4 years worth of photometry over exactly twice the
40.54\,min signal, i.e.\,81.08\,min, results in an average light curve
(Fig.\,\ref{f-fold81}) that unambiguously recreates the morphology
observed in the individual light curves
(Fig.\,\ref{f-lc_braeside_kryoneri} \& \ref{f-lc_ultracam}): a
double-humped modulation with a narrow dip. Dips of such shape are
observed either in strongly magnetic CVs, when the magnetically
funnelled accretion stream passes in front of the white dwarf, or as
grazing eclipses of the hot spot in the accretion disc in non-magnetic
CVs. Stream eclipses in magnetic CVs are relatively unstable features
as the location of the stream within the binary frame changes as a
function of the mass transfer rate and therefore we favour the latter
scenario as the cause of the narrow dips.

The light curve of HS2331 (Fig.\,\ref{f-lc_braeside_kryoneri} \&
\ref{f-fold81}) bears a striking resemblance to that of one of the
most (in)famous and thoroughly studied short period CVs: WZ\,Sge
\citep[see e.g. Fig.\,1 of][]{pattersonetal98-2}.  Both objects
exhibit double humps in their light curves, in both cases one of the
humps appear to suffer from sporadic absorption events, and both
systems show a sharp and relatively shallow eclipse. From the strict
coherence of the 40.54\,min, the detection of periodically recurrent
eclipses, and the close analogy to WZ\,Sge, we
conclude that the orbital period of HS2331 is 81.08\,min, and we
define the following eclipse ephemeris
\begin{equation}
\label{e-ephemeris}
\phi_0 = \mathrm{HJD}\,2451812.67765(35) +  0.05630921(1)\times E
\end{equation}
where phase zero corresponds to the time of mid eclipse. Higher
harmonics of the orbital period are clearly visible in the periodogram
(Fig.\,\ref{f-scarglephot}) at 27.03\,min (53.28\,$\rm d^{-1}$) and
20.27\,min (71.04\,$\rm d^{-1}$).

\subsection{Quiescent superhumps?}
Only a weak signal is detected at the orbital period of 81.08\,min
(Fig\,\ref{f-scarglephot} middle right). This is not too much of a
surprise because of the double-humped structure of the orbital
modulation, which shifts most power to half the orbital
period. Surprising is, however, the detection of a relatively strong
signal close to \Porb, with a period of either $\simeq83.38$\,min
($\simeq17.27\,\rm d^{-1}$) or $\simeq88.51$\,min
($\simeq16.27$\,\id). Both these 1-day aliases are of similar
strength, and their relative power varies between different observing
runs. In contrast to the $35.52\,\id$ orbital signal
(Fig.\,\ref{f-scarglephot} middle left), the periodogram around the
83.38\,min/88.51\,min signals is extremely complex, displaying
significant substructure in excess to the window function, suggesting
either the presence of more than one periodicity and/or a lower degree
of coherence. 
Figure\,\ref{f-fold83} shows the light curve folded over the 83.38\,min
signal detected in the periodogram of Fig.\,\ref{f-scarglephot}
(middle-left panel). The absence of eclipses in this folded light
curve clearly rules out the 83.38\,min signal (and its aliases) being
the orbital period.

The separation of the 83.38\,min signal and the orbital period of
81.08\,min is conspicuously close to those observed for the superhump
period in a large number of short period dwarf nova
\citep[e.g.][]{patterson01-1}. In fact, the period excess found in
HS2331, $\epsilon=(P_\mathrm{sh}-\Porb)/\Porb=0.028$, is similar to
that observed in LL\,And, which has an orbital period close to that of
HS2331. Based on the observational analogies between HS2331 and a
number of well-observed short-period dwarf novae, it is tempting to
ascribe the observed 83.38\,min signal to a permanent superhump. If
this hypothesis were true, HS2331 would be unique, as it displays a
permanent quiescent superhump. If 88.51\,min were the true period its
interpretation in terms of a permanent superhump would result
in $\epsilon=0.092$, much larger than any period excess observed in short
period CVs, and we might as well have discovered a hitherto unknown
phenomenon.

\subsection{Higher frequencies: White dwarf pulsations \& spin?}
As mentioned above, the periodogram of HS2331 contains a number of
distinct signals at high frequencies (Fig.\,\ref{f-scarglephot}), most
pronounced at 5.61\,min (256.6\,\id) and 1.12\,min (1284.7\,\id). We
have inspected in more detail the periodogram computed from the 2003
Kryoneri data, which provides the longest coverage at a high time
resolution, as well as from the ULTRACAM observations, which has the
highest time resolution but covers only
$\sim2.5$\,h. Table\,\ref{t-highfreq} lists the frequencies and
amplitudes of all signals that are unambiguously detected in the two
data sets. The fact that most signals are detected over long periods
of time rules out the CV-typical flickering to explain their origin.

The extremely complex power spectrum in the range around $\sim5$\,min
is indicative of a combination of many frequencies. Such a structure
is reminiscent of the patterns detected in the power spectra of
ZZ\,Ceti stars, non-radially pulsating white dwarfs with a
hydrogen-rich atmosphere. They are found in a narrow range of
effective temperatures, $12\,500\,\mathrm{K}\geq T_{\rm eff}\geq
10\,700$\,K, and show multiperiodic variations with amplitudes
reaching up to $\sim$0.3\,mag and periods ranging from $\sim$100 to
$\sim$1200\,s \citep{clemens93-1}. Resolving the rich frequency
spectrum of ZZ\,Ceti stars related to the various pulsation modes,
their harmonics, and various linear combinations of modes requires
alias-free uninterrupted multi-longitude observing campaigns
\citep[e.g.][]{kepleretal03-1}.  In the case of HS2331, the sampling
of our 2003 Kryoneri run introduces too much of an alias structure
(Fig.\,\ref{f-scarglephot}) to allow a secure identification of all
the various frequencies.

Whereas our short ULTRACAM run is not very useful in identifying the
frequencies present in HS2331, these data show that the power of the
$\sim5$\,min variability increases towards the blue, which is what
would be expected for ZZ\,Ceti pulsations.  Untypical for ZZ\,Ceti
pulsations are the additional large-amplitude signals detected at
extremely high frequencies (Fig.\,\ref{f-scarglephot}). The shortest
period observed in any ZZ\,Ceti star is 70.9\,s in G\,185--32
\citep{castanheiraetal04-1}. Our detection of power at the first
harmonic of the 1.12\,min signal seems to rule out a conventional
ZZ\,Cet pulsation as the origin of the high-frequency signals. 
Yet another stable clock available in a CV is the spin of the white
dwarf, and we tentatively suggest that the 1.12\,min and 0.56\,min
signals are related to the white dwarf spin period and its first
harmonic, respectively. Intriguing is the doublet-like structure of
the 1.12\,min and 0.56\,min signals (Fig.\,\ref{f-scarglephot}). Fast
rotation of the white dwarf will result in extremely non-linear
perturbations of the pulsation frequencies, and may be responsible for
this structure. Alternatively, the doublet-structure may reflect a
sideband frequency, as typically observed in intermediate polars. The
choice of the lower frequency alias in the doublets is somewhat
ambiguous: assuming that 1277.7\,\id\ and 2555.5\,\id\ are the true
frequencies of the doublet and its first harmonic, the implied beat
frequency is 7\,\id, which is nearly identical to the spectroscopic
period found from the radial velocity variations of the Balmer
emission line wings (Sect.\,\ref{s-ana_spec}).

\begin{table}
\caption[]{\label{t-highfreq}High-frequencies signals
detected in the photometry of HS\,2331+3905.}
\newcommand{\ali}{\FN{\dag}}
\begin{flushleft}
\begin{tabular}{rrrrrr}
\hline\noalign{\smallskip}
\multicolumn{2}{c}{Kryoneri} & \multicolumn{4}{c}{WHT} \\
\multicolumn{1}{c}{$f$} & \multicolumn{1}{c}{clear} &
\multicolumn{1}{c}{$f$} & \multicolumn{1}{c}{$u'$} & 
\multicolumn{1}{c}{$g'$} & \multicolumn{1}{c}{$r'$} \\
\multicolumn{1}{c}{$[\id]$} & \multicolumn{1}{c}{[mmag]} &
\multicolumn{1}{c}{$[\id]$} & \multicolumn{1}{c}{[mmag]} & 
\multicolumn{1}{c}{[mmag]} & 
\multicolumn{1}{c}{[mmag]} \\
\noalign{\smallskip}\hline\noalign{\smallskip}
205.7      & 3.6  & 206  &  9.6 &  6.7 &  7.4 \\
212.5      & 4.5  & 216  & 15.0 & 10.1 & 11.2 \\
213.2      & 4.8  & \\
           &      & 226  & 14.2 & 11.9 & 12.1 \\
241.2      & 6.4  & \\
256.6      & 10.0 & 266  & 36.3 & 20.2 & 22.1 \\
278.6      & 4.9  & \\
283.0      & 3.6  & 288  & 11.7 &  7.7 &  8.6 \\
           &      & 474  &  6.6 &  5.3 &  4.8 \\
           &      & 535  &  7.2 &  6.6 &  6.5 \\ 
1029.1     & 2.6  & 1021 &  8.5 &  6.8 &  6.9 \\
1276.7     & 7.0  & \\
1284.7     & 15.8 & 1283 & 35.1 & 20.7 & 20.6 \\
2554.5\ali & 6.2  & 2554 &  7.6 &  9.4 &  6.2 \\
2569.5     & 3.7  & 2570 & 11.0 &  4.4 &  6.2 \\
\noalign{\smallskip}\hline\noalign{\smallskip}
\end{tabular}

\fn{\dag} The 1-day aliases at 2555.5 has nearly equal power.
\end{flushleft}
\end{table}

\section{Analysis: Spectroscopy}
\label{s-ana_spec}
\subsection{Radial velocities}
Radial velocities were measured using the convolution technique first
outlined by \citet {schneider+young80-2} and developed by \citet
{Shafter83-1}.

The emission line profile is convolved with two Gaussians of equal
FWHM (of the order of the wavelength resolution) but with opposite
signs.  The peaks of the Gaussians are separated by a distance chosen
in order to measure as far out as possible into the wings, while still
retaining an adequate signal-to-noise level. By doing this we ensure
that the measured radial velocities trace the inner part of the
accretion disc, where one may hope to find an azimuthally symmetric
structure that traces the orbital motion of the white dwarf. Prior to
the actual radial velocity measurement the spectra were transformed
into the heliocentric rest frame.


\subsubsection{The end of a myth...}
Radial velocity variations were measured for each individual observing
run (Table\,\ref{t-obslog}) using the strongest Balmer lines as well
as He\,I $\lambda5875$ and He\,I $4471$. In all cases the Gaussians
separation was set to values between 2600 $\rm km\,s^{-1}$ and 3000
$\rm km\,s^{-1}$. The radial velocity measurements were then subjected
to the Scargle period analysis in the \texttt{MIDAS/TSA} context. The
results for each individual run did not depend on the emission line
used for the analysis, and we will discuss in the following the radial
velocity variations measured in \Hg\ (Calar Alto 2000; we opted
against \Hb\ because of the potential contamination from the nearby He
lines) and \Ha\ for all the other runs.

\begin{figure*}
\centerline{\includegraphics[angle=0,width=16cm]{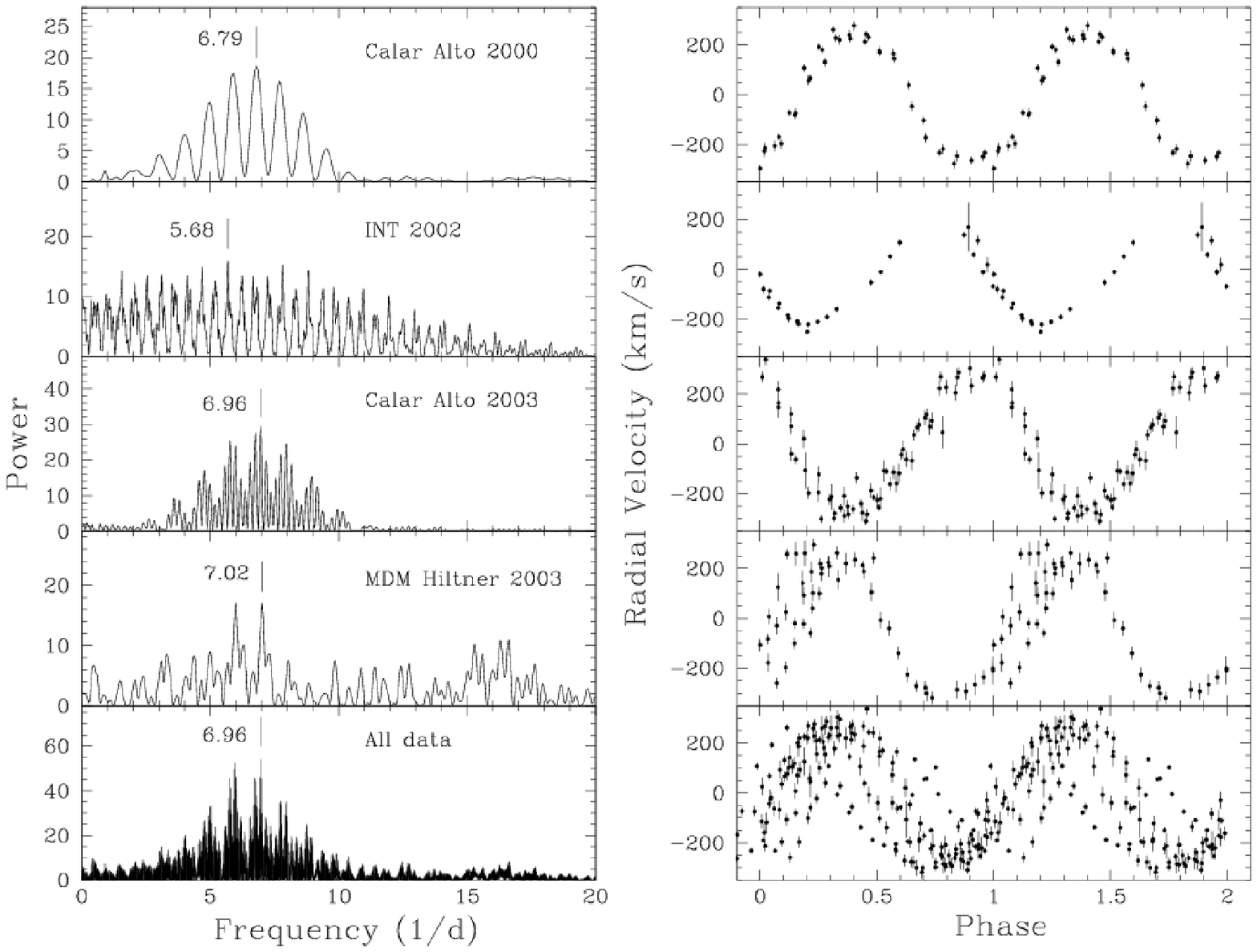}}
\caption[]{\label {f-scarglespec} {\it Left Panel:} Scargle
periodograms of the individual data sets (see Table\,\ref{t-obslog})
and of all the data combined together. {\it Right Panel:} Radial
velocity curves of the individual data sets and of all the data sets
combined together, folded over the highest peak in their respective
Scargle periodograms (shown on the left). The zero phase is taken to
be the first point in each of the data sets respectively and the
curves are repeated twice for clarity.}
\end{figure*}

\begin{figure*}
\includegraphics[angle=-90,width=8.8cm]{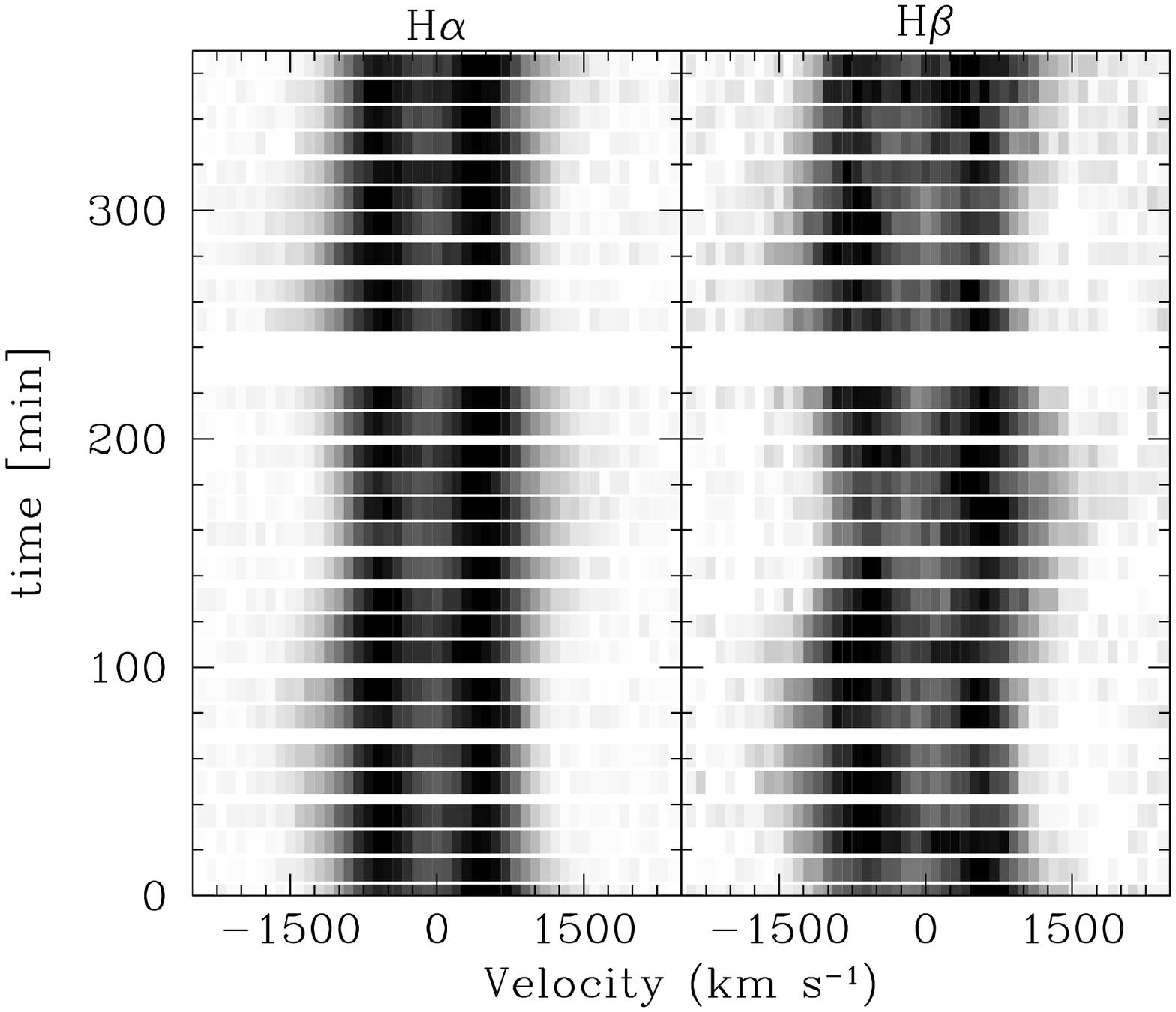}
\includegraphics[angle=-90,width=8.8cm]{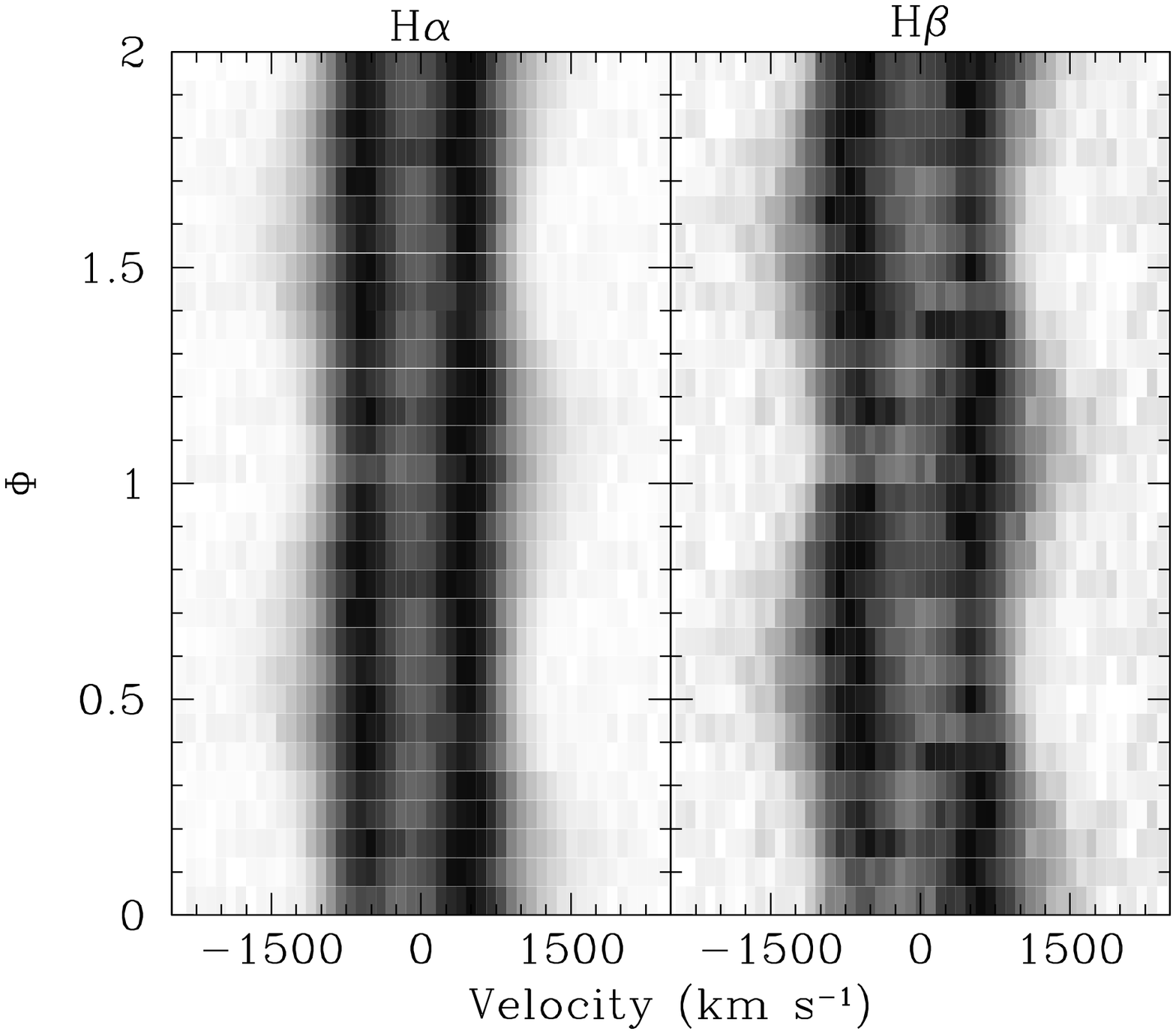}

\caption[]{\label {f-trail} {\it From left to right:} Trail spectra of
  \Ha\ and \Hb\ computed from the largest data set within an observing
  night (i.e. Calar Alto 2003, Sep\,24; see Table\,\ref{t-obslog}), and
  the same \Ha\ and \Hb\  trail spectra but this time folded over the
  orbital period (i.e. $P_{\rm orb} = 81.08$\,min). The white stripes
  in the trail spectra of the left panel are caused by lack of
  observations during the corresponding times.}
\end{figure*}

Figure\,\ref{f-scarglespec} shows the periodograms and the
corresponding radial velocity curves folded over the highest peaks
detected in each of the data sets. The bottom panel shows the
periodogram and folded radial velocity curve obtained when combining
all data. The Calar Alto 2000 data show a smooth sinusoidal radial
velocity variation with an amplitude of $\sim220$\,\kms\ and a period
of 3.53\,h, \textit{which is not anywhere close to the orbital period
determined from the eclipses!} This makes HS2331, to our knowledge,
the first CV in which the dominant radial velocity variation is not
modulated on the orbital period. With HS\,2331, there are now
four CVs in which the photometric modulation and the spectroscopic
modulation are significantly different; the other three being FS\,Aur,
GW\,Lib and Aqr1 \citep[see Table\,2 of][]{woudtetal04-1}. The origin
of the second period in these systems is much of a mystery as the
$\sim3.5$\,h periodicity in HS\,2331, but contrary to HS\,2331, in
FS\,Aur, GW\,Lib and Aqr1 the spectroscopic variation is believed to
be orbital. Fig.\,\ref{f-trail} shows the trailed spectra of \Ha\ and
\Hb\ obtained from the observing night with the largest data set
(i.e. Calar Alto 2003, Sep 11; Table\,\ref{t-obslog}) in minutes and
phase folded with the orbital period respectively. The 3.45\,h period
is clearly seen in the wings of the Balmer emission lines.

The analysis of the other runs reveals that the radial velocity
variation changes on time scales of several days. The structure of the
periodograms becomes increasingly complicated with the length of the
run, and folding the radial velocity measurements over the strongest
peak (or any other peak) will not provide a smooth radial velocity
variation (Fig.\,\ref{f-scarglespec}). The strongest signals in the
periodograms are found at 4.22\,h (INT\,2002), 3.45\,h (Calar Alto
2003), 3.42\,h (MDM Hiltner 2003) and 3.45\,h (all data). All runs,
with the exception of the INT data, give very close period values
which appear consistent with each other. Close inspection of the
bottom panel of Fig.\,\ref{f-scarglespec} shows that, when folded over
the $\sim3.45$\,h signal, (the highest peak detected in the
periodogram calculated from combining all the spectroscopic data
together) a shift between the individual nights appears (Calar Alto
2000/2003 and MDM 2003 data). In the case of the INT observing run,
the poor sampling of the data (Table\,\ref{t-obslog}) may prevent us
from detecting the true underlying periodicity at this epoch and the
relatively smooth appearance of the radial velocity curve folded over
4.22\,h (the highest peak in the periodogram) may be
coincidental. Statistically, there is no strong preference for any of
the aliases found in the range $\sim 4\,\rm d^{-1}-10\,d^{-1}$. 

We conclude that a persistent large-amplitude radial velocity
variation with a period of $\sim$3.5\,h is present in HS\,2331,
however, this variation is not coherent and its period and/or phase
drifts on time scales of days.

\subsubsection{\label{s-rvdetrended}...sanity refound}
\begin{figure*}
\centering
\includegraphics[angle=270,width=18cm]{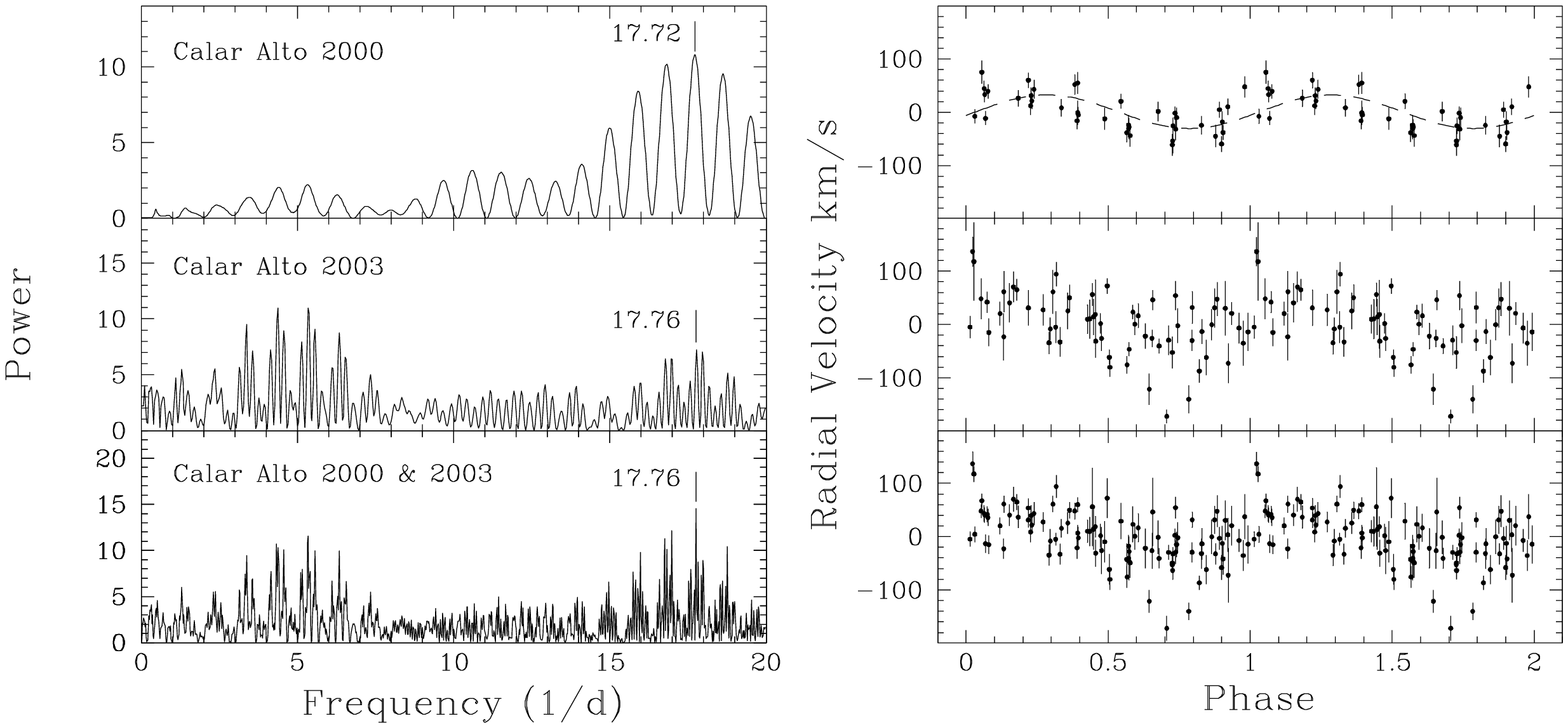}
\caption[]{\label{f-detrended_all} {\it Left Panel:} From top to
  bottom, Scargle periodograms of the detrended Calar Alto 2000 and
  Calar Alto 2003 spectroscopic data sets and a combination of both
  sets (i.e. Calar Alto 2000 \& Calar Alto 2003). The radial velocity
  measurements were detrended by subtracting a sine wave with their
  respective highest peak value (see the left panel of
  Fig.\,\ref{f-scarglespec}) as the frequency. The combined Calar Alto
  2000 \& 2003 Scargle periodogram in the bottom panel was computed
  after detrending the data sets individually. {\it Right panel:}
  Detrended radial velocity curves folded over the orbital period
  obtained from the photometry in Sec.\,\ref{s-ana_photo} (i.e.
  $17.76\,\rm{d^{-1}}$ or 81.08\,min). The zero phase in these curves
  corresponds to the photometric zero phase as defined in
  Sec.\,\ref{s-ana_photo} and the curves are repeated twice for
  clarity. Also shown on the top right panel is the best fit to the
  detrended radial velocity curve (see text for more details).}
\end{figure*}

The fact that the dominant radial velocity variation in HS2331 does
not occur on the orbital period is most surprising, and alarming, as
radial velocity variations measured from the emission lines are
routinely used not only to determine the orbital periods of CVs, but
also to estimate the radial velocity of the white dwarf.  In order to
probe for lower amplitude radial velocity variations, we have
detrended the Calar Alto 2000 and 2003 radial velocity measurements
with a sine wave adopting the frequency of the strongest peak in the
corresponding periodogram, 3.53\,h and 3.45\,h, respectively. The
Scargle periodograms of the detrended data
(Fig.\,\ref{f-detrended_all}; left panel) contain significant power at
frequencies which are consistent with the orbital period derived from
the photometry (Sect.\,\ref{s-ana_photo}): $81.26\pm0.96$\,min (Calar
Alto 2000), $81.08\pm0.18$\,min (Calar Alto 2003), and $81.08\pm0.04$\,min
(combining both sets). Whereas the signal of the orbital period
completely dominates the periodogram of the  2000 data, significant
residual power is left near 3.45\,h in the case of the 2003 data. The
main difference between the two data sets is the extent of the
observations. While the 2000 data were obtained during two consecutive
nights, the 2003 data was obtained during three nights spanning a total of
six nights. As the dominant radial velocity variation drifts on times
scales of a few days, detrending with a sine with a fixed period will
become increasingly imperfect with increasing length of the observing
run. Hence, the residual low-frequency power in the case of the 2003
data reflects the imperfection of the applied detrending.

Folding the detrended radial velocities over the orbital period
(i.e. 81.08\,min) results in a sinusoidal radial velocity curve
(Fig.\,\ref{f-detrended_all}; right panel), which is smoother for the
2000 data than for the 2003 data, probably again a result of the
cleaner detrending possible for these observations.  The consistent
detection of the 81.08\,min period both, in the photometry and in the
spectral radial velocity variations lends further support for our
conclusion that this is indeed the orbital period of HS\,2331. 

A sine fit to the folded detrended radial velocities of the Calar Alto
2000 data (top right panel of Fig.\,\ref{f-detrended_all}) results in
an amplitude of $\simeq32\,\mathrm{km\,s^{-1}}$, a systemic velocity
of $\sim1$\,\kms, and a phase offset with respect to mid eclipse of
$\sim 13^{\circ}$.  The amplitude, interpreted as radial velocity of
the white dwarf \Kwd\, as it is commonly done, is entirely consistent
with what is observed in short-period CVs
\citep[e.g.][]{thorstensen+fenton03-1}. The phase offset with respect
to the expected motion of the white dwarf is most likely related to
the fact that the eclipse is caused by an obscuration of the hot spot
in the accretion disc, and not of the white dwarf itself.  In addition,
even in ``well-behaved'' CVs the broad wings of the emission lines do
not necessarily exactly trace orbit of the white dwarf, most likely
because of slight asymmetries in the disc.

\subsection{\label{s-sed}A model for the spectral energy distribution}
\begin{figure*}
\includegraphics[angle=270,width=\textwidth]{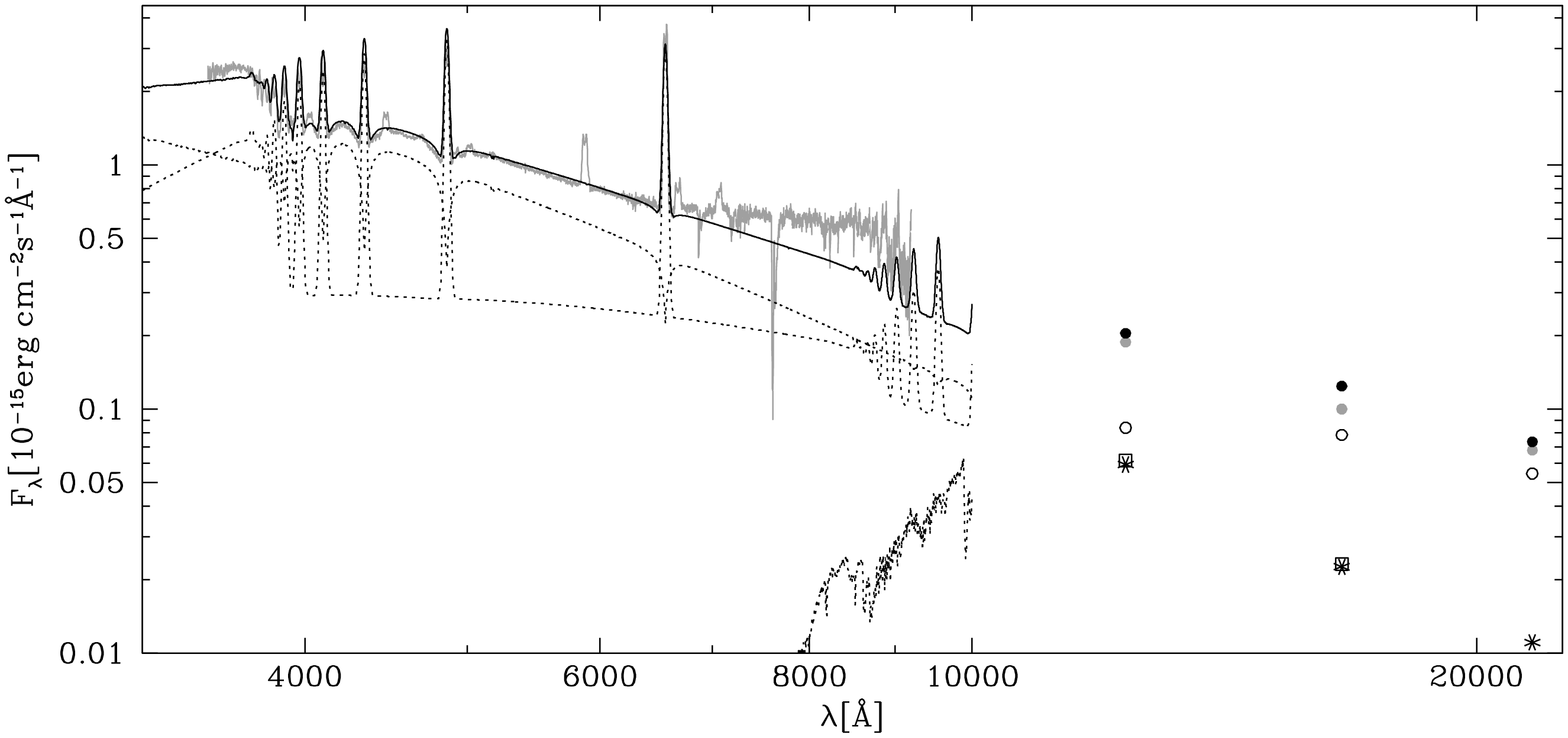}
\caption[]{\label{f-sed_optir} Three component model to the optical-IR
spectral energy distribution of HS2331. Plotted in gray are the
observed average spectrum from Fig.\,\ref{f-average} along with the
2MASS $J$, $H$, and $K$ fluxes (filled circles). Plotted in dotted
lines are the white dwarf model (open squares denote the IR fluxes), the
isothermal slab model representing the optically thin disc emission
(stars denote the IR fluxes), and the spectral contribution of a L2
donor star (open circles denote the IR fluxes). The sum of the three
components is shown as black solid line and as black filled circles.}
\end{figure*}

We have combined our optical data with the 2MASS $J$, $H$, and $K$
magnitudes of HS2331 (Fig.\,\ref{f-sed_optir}), and modeled the
optical/IR spectral energy distribution (SED) with a three-component
model comprising: (1) a synthetic white dwarf spectrum to account for
the observed broad Balmer absorption lines. A grid of model spectra
covering $\Teff=8\,000-15\,000$\,K was calculated with TLUSTY and
SYNSPEC (\citealt{hubeny88-1, hubeny+lanz95-1}). We fixed the surface
gravity to $\log g = 8.0$, corresponding to $\Mwd\simeq0.6\,\Msun$,
and the photospheric abundances to 0.1 of their solar values. (2) The
emission of an isothermal and isobaric slab with a finite depth to
account for the observed Balmer emission lines and any associated
continuum. This model is described in detail in
\citet{gaensickeetal97-1} and \citet{gaensickeetal99-1}. In brief, the
free parameters for the generation of this component were the
temperature of slab and the column density of the slab along the line
of sight.  The temperature accounts for the ionisation/excitation of
the lines and the optically thick blackbody envelope, and the column
density determines the ratio in optical depth between continuum and
lines, i.e. the equivalent width of the lines. (3) A late-type
spectral template to account for the emission of the secondary star in
HS2331. The library of templates covered spectral types M0.5 to M9
from \citet{beuermannetal98-1} and L0 to L8 from
\citet{kirkpatricketal99-1} and \citet{kirkpatricketal00-1}.  The
L-dwarf type stars were normalized to surface fluxes using the radii
and distances given by \citet{dahnetal02-1}, the M-dwarf templates
provided by Beuermann were already normalized in this way.

\begin{figure}
\includegraphics[angle=270,width=\columnwidth]{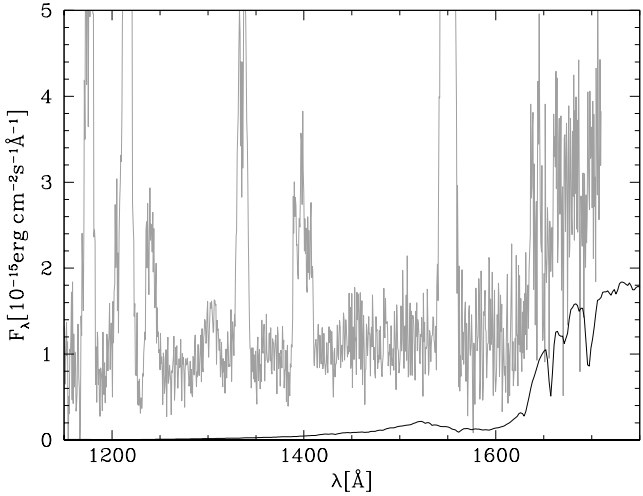}
\caption[]{\label{f-stis_fit} Gray line: the STIS FUV spectrum of
  HS2331 from Fig.\,\ref{f-hst}. Black line: the white dwarf model
  from Fig.\,\ref{f-sed_optir} extended into the STIS range. The
  spectrum of a $\sim10\,500$\,K white dwarf reproduces well the
  sudden increase of flux observed longwards of 1600\,\AA.}
\end{figure}

We started by approximately modeling the observed broad Balmer
absorption lines with a white dwarf model, which suggested
$\Teff\sim10\,000$\,K and a white dwarf flux contribution of $\sim
80$\% at 5000\,\AA\ (Fig.\,\ref{f-sed_optir}). The next step was to
extrapolate the white dwarf model established in this way into the FUV
wavelength range covered by STIS
(Fig.\,\ref{f-stis_fit}). Interpreting the sudden flux up-turn at
wavelengths $\lambda\ga1650$\,\AA\ as photospheric emission from the
white dwarf we refine the white dwarf temperature to
$\Teff\simeq10\,500\pm250$\,K. The red end of the STIS range is very
sensitive to the white dwarf temperature because of the strong
temperature dependence of the quasimolecular $H_2$ absorption at
1600\,\AA. The flux scaling factor of our white dwarf model implies a
distance of $\sim90$\,pc. The systematic error in this distance
estimate depends primarily on the assumed white dwarf mass, assuming a
mass in the range $0.4-0.9$\,\Msun\ results in $d=90\pm15$\,pc.  A
low distance to HS2331 is supported by the large proper motion of the
star, $\mu\simeq0.14\arcsec/\mathrm{yr^{-1}}$, which has been derived
from the comparison of the DSS1 and DSS2 images of the system. An
isothermal/isobaric slab with a temperature of $T=6500$\,K and a
surface density of $\Sigma=1.8\times10^{-2}\,\rm g\,cm^{-2}$ accounts
well for the observed Balmer emission and the Balmer jump, and
contributes $\sim20$\,\% of the flux at 5000\,\AA. Assuming $d=90$\,pc
the area of this simple ``disc'' component is comparable to the size
of the white dwarf Roche lobe. The red end of our optical spectrum
shows no evidence for the TiO absorption bands that are observed in
CVs where the late-type donor star significantly contributes to the
red/near-IR spectrum. In the last step of our SED modeling we add the
spectral contribution of the secondary, assuming $R_2=8\times10^9$\,cm
(corresponding to $\Porb=81.08$\,min and $q=\Mwd/\Msec=9$; see
Sect.\,\ref{s-evostate}). The flux limit imposed by the 2MASS $J$,
$H$, and $K$ measurements constrain the spectral type of the secondary
to be $\sim$L2 or later (Fig.\,\ref{f-sed_optir}).

Whereas our three-component model of the FUV-to-IR SED of HS2331 is
overall rather satisfying, two regions of excess flux are observed:
the FUV continuum below $\sim1600$\,\AA\ significantly exceeds the
predicted white dwarf flux (the simple ``disc'' model and the
secondary contribute no flux in the FUV), and the optical continuum in
the range $\sim7000-9000$\,\AA\ exceeds the flux of the sum of all
three components. We believe that the most likely explanation for this
excess flux is that our model for the disc emission is too naive~--~we
just assume a single temperature/density pure hydrogen slab. The FUV
emission lines clearly indicate that the disc contains hotter regions,
which will contribute to some extent to the FUV continuum. The
excess flux observed at the red end of our optical spectrum may be
related to the Paschen continuum of a somewhat colder part of the
disc. It is likely that a multi-temperature/density model of the
quiescent disc could account for the observed excess flux, however, we
feel that the number of additional free parameters does not warrant
this exercise. 

\section{Discussion}
\label{s-discussion}

Before discussing the nature of the system, we briefly summarise the
results of our analysis.

\begin{enumerate}
\item HS2331 has been identified as a CV because of its emission lines
detected in the HQS.
\item Coherent low-amplitude eclipses detected over four years
unequivocally determine the orbital period to be
$\Porb=81.08$\,min. The mean light curve of HS2331 bears extreme
similarity to that of WZ\,Sge.
\item The optical and FUV spectrum of HS2331 is dominated by the
photospheric emission of a white dwarf with $\Teff\simeq10\,500$\,K.
\item The power spectrum of our optical photometry contains a number
of significant signals in the range $5-6$\,min, which we interpret
as non-radial ZZ\,Ceti pulsations. 
\item Additional signals are found at periods of 1.12\,min and
  0.56\,min, which we tentatively identify as the spin period of the
  white dwarf and its first harmonic. 
\item The distance of the system is $d\simeq90\pm15$\,pc, derived from
  modelling the emission of the white dwarf. This low distance is
  consistent with the high proper motion of the system,
  $\mu\simeq0.14\arcsec/\mathrm{yr^-1}$.
\item Combining our optical spectroscopy of HS2331 with the 2MASS $J$, $H$, and
  $K$ magnitudes of the system, and adopting $d=90$\,pc and a mass
  ratio of  $q=\Mwd/\Msec=9$ (see below, Sect.\,\ref{s-evostate}), we
  find that the donor star must be of extreme late spectral type, L2
  or later.
\item Our extensive optical photometry consistently contains a strong
  signal at a period somewhat longer than \Porb. Selecting the alias of that
  pattern which is closest to \Porb\ gives 83.38\,min, which is
  2.8\,\% longer than the orbital period. 
\item Our spectroscopic data set reveals consistently the presence of
  large-amplitude radial  velocity variation with a period of
  $\sim3.5$\,h. However, this radial velocity  modulation is not
  strictly coherent, but appears to drift in either period or phase
  on a time scale of days.
\item After detrending the radial velocity measurements with a 3.5\,h
  sinusoid, a radial velocity variation with a period of
  $81.08\pm0.04$, i.e. \Porb, is detected with an amplitude of
  $\simeq32$\,\kms.
\end{enumerate}

\subsection{\label{s-evostate} The evolutionary state of HS\,2331+3905}
A number of arguments suggest that HS2331 is a CV in a late stage of
its evolution. Its orbital period is close to the observed period
minimum of CVs.  A very low mass transfer rate is indicated by the low
temperature of the white dwarf, which is a measure of the secular mean
accretion rate \citep{townsley+bildsten03-1}. This is consistent with
the absence of observed outbursts. All our detailed photometric and
spectroscopic campaigns carried out since September 2000 have found
HS2331 near $V\simeq16.5$. The CCD-equipped 0.37\,m Rigel Telescope of
the University of Iowa has been used to monitor HS2331 on 16 nights in
October to December 2003 and on 15 nights in May/June 2004, finding
the system at an instrumental magnitude of $16.4\pm0.1$. Additional
visual monitoring of HS2331 has been carried out during $\sim250$
nights since March 2003 using a 8" Schmidt-Cassegrain telescope, with
no gaps in the coverage being longer than one week. On no occasion has
the star been detected, with a limiting magnitude of $\sim13.5\pm0.5$.
In many aspects, HS2331 resembles the well-studied short-period dwarf
nova WZ\,Sge, which shows extremely bright superoutbursts every 2--3
decades. If this analogy were correct, we would predict an outburst
magnitude for HS2331 of $V\simeq9-10$.

Yet another indication of HS2331 being an old evolved CV is the very
late spectral type of the secondary star. Assuming a white dwarf mass
of 0.6\,\Msun\ and a secondary mass of $0.07$\,\Msun\ (as it may be
realistic for \Msec\ near the minimum period), i.e. $q\simeq9$, and a
distance of 90\,pc, our model for the SED of HS2331 suggests a
spectral type later than L2 (Sect.\,\ref{s-sed}), which makes this
system one of the best candidates for a brown dwarf donor. Detailed
infrared spectroscopy will be required to test this hypothesis.

Using the same stellar mass assumption and an inclination of
$\simeq75^{\circ}$, based on the detection of grazing eclipses in the
optical light curve, the radial velocity variation of the white dwarf
is expected to be $K_1\simeq36$\,\kms, which is consistent with the
amplitude of the 81.08\,min radial velocity variation detected in the
spectroscopy, after detrending the data for the dominant 3.5\,h
modulation (Sect.\,\ref{s-rvdetrended}).

\subsection{A permanent superhumper?}
From our four years of photometric data, HS\,2331 appears to be a
permanent superhumper with a likely superhump period of $P_{\rm
sh}=83.38$\,min, i.e. a period excess
$(P_\mathrm{sh}-\Porb)/\Porb=2.8\%$. This finding is exceptional, as
superhumps are normally detected during superoutbursts of SU\,UMa type
dwarf novae or in some nova-like variables during their high states,
i.e. always during phases of high mass transfer in the accretion
disc. Superhumps are thought to be related to the precession of
tidally unstable eccentric accretion discs in close binaries with
extreme mass ratios \citep[e.g.][]{osaki85-1, murray98-1}. 

If our interpretation of the photometric signal is correct, and indeed
$P_\mathrm{sh}=83.38$\,min, the implied period excess of 2.8\% is not
particularly small \citep{patterson01-1}, somewhat arguing against an
extreme mass ratio in HS2331. However, considering the unique nature
of HS2331, it is not clear if Patterson's relation can be directly
adopted for this system. 

The strangest and most disconcerting observational fact about HS2331
is without doubt the detection of a dominant radial velocity
variability with a period of 3.5\,h, which does not correspond to the
orbital motion of the system. This periodicity resulted from measuring
changes in the wings of the Balmer emission lines and therefore reflects
variability in the inner disc region of the systems, however we have at
present no explanation for the physical origin/significance of its
3.5\,h period. Again, in a very speculative approach, we suggest that
the period (or phase) drift of this radial velocity variation may be
related to the presence of permanent superhumps. The precession period
of the accretion disc implied by the orbital period and the superhump
period is $\sim2$\,d, which is comparable to the time scale on which
the 3.5\,h radial velocity variation varies. An intensive
spectroscopic campaign covering several of the putative precession
cycles could test this hypothesis.

Another important conclusion from the radial velocity analysis is that
care should be taken when implicitely assuming that the dominant
radial velocity variation detected in cataclysmic variables (and
related objects) reflects the orbital period. Whereas this is contrary
to the customary approach for the past decades, the results obtained
for HS\,2331 are a clear warning that this custom could occasionally
lead to wrong interpretations.

\subsection{The brightest CV white dwarf pulsator}
Until recently, only a single CV containing a white dwarf pulsator has
been discovered: GW\,Lib \citep{vanzyletal00-1,
vanzyletal04-1}. GW\,Lib is only partially understood, due to the lack
of a proper mode identification of the observed pulsation frequencies,
and more severely, due to the circumstance that the temperature of its
white dwarf, $T_{\rm eff}=14\,700$\,K, is actually significantly above
the blue end of the ZZ\,Ceti instability strip
\citep{szkodyetal02-4}. Three additional CVs containing pulsating
accreting white dwarfs have been discovered in the SDSS:
SDSS\,J013132.39--090122.3, SDSS\,J161033.64--010223.3, and
SDSS\,J220553.98+115553.7 \citep{warner+woudt03-1, woudt+warner04-1}.

We have shown that HS2331 exhibits the photometric behaviour typical
of ZZ\,Ceti pulsators, showing a multiperiodic and complex power
spectrum in the range $\sim 60$\,s to $\sim300$\,s. With
$V\simeq16.5$, HS2331 is by far the brightest CV white dwarf pulsator,
making it the prime target for detailed follow-up studies.  The white
dwarf temperature implied by the SED of HS2331
($\Teff\simeq10\,500$\,K, assuming $\Mwd\simeq0.6$\,\Msun,
Sect.\,\ref{s-sed}) is just slightly below the red end of the ZZ\,Ceti
instability strip for single white dwarfs
\citep[e.g.][]{bergeronetal04-1}. The ultra-short periods detected in
HS2331, with periods of 1.12\,min and 0.56\,min which are too small
for conventional $g$-mode pulsations, remains a puzzle. One possible
solution is to assume that 1.12\,min is the white dwarf spin period,
and 0.56\,min its first harmonic. Polarimetry and X-ray detection of
the 1.12\,min period would be the ultimately proof of this hypothesis.

A more detailed understanding of the white dwarf pulsations in CVs is
just at its beginning, as the first models for the driving mechanism
are being developed \citep{townsletetal04-1}, and a larger sample of
these objects is discovered. Future potential  
asteroseismological applications in CV white dwarfs include measuring the
white dwarf masses, as well as studying the accreted envelope.

\section{Conclusion}
We have discovered a new short orbital period CV, HS\,2331, which
displays an overwhelming  wealth of observable phenomena, and we
encourage detailed follow-up studies over all wavelength ranges and
time scales to exploit the opportunities that this system offers for
an improved understanding of CV evolution and accretion onto white
dwarfs. 

\acknowledgements We thank the referee, Patrick Woudt, for a prompt
and helpful report. SAB acknowledges support from NASA through grants
GO-9357 and GO-9724 from the Space Telescope Science Institute, which
is operated by AURA, Inc., under NASA contract NAS5-26555. BTG was
supported by a PPARC Advanced Fellowship, The HQS was supported by the
Deutsche Forschungsgemeinschaft through grants Re\,353/11 and
Re\,353/22. We are very grateful to Klaus Beuermann and Davy
Kirkpatrick for giving us access to their spectral templates for M-
and L-dwarfs. We would also like to thank Pablo Rodriguez-Gil,
Matthias Schreiber, Klaus Beuermann, Luisa Morales, Dean Townsley,
Lars Bildsten and Phil Arras for helpful discussions and to Spiridon
Kitsionas for a careful reading of the manuscript. P.\,Schmeer is very
grateful to Robert Mutel and his students for scheduling some of the
CCD monitoring of HS\,2331. We thank the director of the Calar Alto
observatory for the generous allocation of discretionary time for this
project.

\bibliographystyle{aa}

\end{document}